\documentclass[aip,reprint,pop]{revtex4-1}

\usepackage{graphicx}
\usepackage[tight]{subfigure}
\usepackage{epstopdf}
\usepackage{amsmath}
\usepackage{dcolumn}
\usepackage{bm}
\usepackage{upgreek}
\usepackage[colorlinks=true, linkcolor=blue]{hyperref}

\begin{document}

\title{Centrifugal instability in the regime of fast rotation}
\author{R. Gueroult}
\affiliation{Laplace, Universit\'{e} de Toulouse, CNRS, INPT, UPS, 31062 Toulouse, France}
\author{J.~M.~Rax}
\affiliation{Universit\'{e} de Paris XI - Ecole Polytechnique, LOA-ENSTA-CNRS, 91128 Palaiseau, France}
\author{N. J. Fisch}
\affiliation{Princeton Plasma Physics Laboratory, Princeton University, Princeton, NJ 08543, USA}

\date{\today}

\begin{abstract}
Centrifugal instability, which stems from a difference between the azimuthal angular drift velocity of ions and electrons, is studied in the limit of fast rotation for which ions can rotate up to twice as fast as electrons. As the angular velocity approaches the so-called Brillouin limit, the growth rate for the centrifugal instability in a collisionless solid-body rotating plasma increases markedly, and is proportional to the azimuthal mode number. For large wavenumbers, electron inertia effects set in and lead to a cut-off. Interestingly, conditions for the onset of this instability appear to overlap with the operating conditions envisioned for plasma mass separation devices.
\end{abstract}

\maketitle

\section{Introduction}

Crossed field or $\mathbf{E} \times \mathbf{B}$ configurations, where an electric field $\mathbf{E}$ exists perpendicularly to the magnetic field $\mathbf{B}$, are found across a large variety of environments including space physics~\cite{Tsuda1969}, tokamaks edge region~\cite{Taylor1989} and laboratory plasma sources~\cite{Abolmasov2012}.  A characteristic feature of crossed field configurations is the associated $\mathbf{E} \times \mathbf{B}$ drift. In laboratory plasmas, this drift can be exploited to fulfil a particular role by adequately tailoring the field topology and strength. For example, the electron $\mathbf{E} \times \mathbf{B}$ drift is key in the efficiency of Hall thrusters~\cite{Morozov2000,Boeuf2017}, Penning plasma sources and magnetron discharges~\cite{Abolmasov2012}. Another application of crossed field configurations is plasma rotation control, where the field orientation is chosen so that charged particles drift azimuthally.  Plasma rotation holds promise for mass separation applications (see, \emph{e.~g.}, Refs~\cite{Bonnevier1966,Lehnert1971,Prasad1987,Ohkawa2002,Fetterman2011,Gueroult2014,Rax2016}), and has been shown to yield unusual heat capacity effects~\cite{Geyko2017}. In addition, plasma rotation has recently been suggested to compensate the vertical drift associated with toroidal magnetic field in a torus~\cite{Rax2017}. However, experiments often reveal that perpendicular transport in crossed field configurations exceeds significantly classical predictions, which in turn impedes the performances of these devices. This discrepancy is generally believed to stem from instabilities and turbulence~\cite{Redhead1988,Horton1999,Litvak2001,Litvak2004,Smolyakov2017}.

Simon~\cite{Simon1963} and Hoh~\cite{Hoh1963} independently demonstrated that a magnetized plasma with transverse electric field can become unstable in the presence of ion-neutral collisions. In what is now known as the Simon-Hoh instability (SHI), both ions and electrons are magnetized. The SHI stems from a difference between ions and electrons drift velocity owing to the larger effect of neutrals on ions. Ions are said to \emph{drag} behind electrons. This leads to charge separation and the subsequent formation of an electric field along the $\mathbf{E}\times \mathbf{B}$ direction. For certain density profiles, this perturbed electric field can cause an initial density perturbation to grow. Simon and Hoh showed that this configuration is unstable if $\bm{\nabla}\phi\cdot\bm{\nabla} n<0$, where $\phi$ and $n$ are the equilibrium plasma potential and density, respectively. 




A variation on the SHI has been found to occur in collisionless plasmas where electrons are magnetized but ions are not~\cite{Sakawa1993}. Although it stems in this regime from the weak magnetization of ions and not from collisions, this configuration also leads to a difference in drift velocity along the $\mathbf{E}\times \mathbf{B}$ direction. Similarly to the SHI, this configuration is unstable for $\bm{\nabla}\phi\cdot\bm{\nabla} n<0$~\cite{Sakawa1992,Sakawa1993,Sakawa2000}. Strictly speaking, this instability requires a drift velocity larger than the square of the ion sound speed divided by four times the electron diamagnetic velocity~\cite{Frias2012,Frias2013,Smolyakov2017}. 
Because it shares the same instability mechanism as the SHI, namely a difference in drift velocity between ions and electrons, this instability is referred to as the \emph{modified} Simon-Hoh instability (MHSI). Note however that the notion of drift velocity for ions is here  ill defined since ions are unmagnetised. The MHSI is more generally referred to as an \emph{anti-drift} mode~\cite{Fridman1964,Frias2013,Smolyakov2017} because the real frequency of this mode is inversely proportional to the diamagnetic frequency.

A third mechanism leading to a difference between ion and electron drift velocity is centrifugal forces in a fully-ionized magnetized plasma column rotating under a radial electric field. This instability mechanism was first uncovered by Rosenbluth \emph{et al.}~\cite{Rosenbluth1962} for a slowly rotating plasmas, and is a particular case of gravitational or \emph{flute} instability~\cite{Rosenbluth1957}. Chen then confirmed and extended these results using a two-fluid model~\cite{Chen1966}. Following these early contributions, the influence on stability of a variety of parameters including boundary conditions~\cite{Chen1967,Rognlien1973}, electric field non-uniformity~\cite{Rosenbluth1965,Perkins1971,Jassby1972,Rognlien1973} and plasma density~\cite{Simon1966} was studied. In contrast, the regime of fast plasma rotation appears to have received limited attention. Although Chen considered this question~\cite{Chen1966}, his results are limited to regimes where $|\varpi_i |/\Omega_i\ll 1$, with $\varpi_i$ the ion equilibrium angular frequency and $\Omega_i$ the ion cyclotron frequency. Also, we note that centrifugal instabilities with $|\varpi |/\Omega_i\sim \mathcal{O}(1)$ were more recently studied in a rotating vacuum arc centrifuge~\cite{Hole2002}, but the radial electric field is in that case negative~\cite{Ilic1973}

Since centrifugal instabilities stem from differences in azimuthal drift velocity, they can occur both with positive and negative radial electric field~\cite{Chen1966}. Nonetheless, the configuration is asymmetrical since centrifugal forces speed up ion rotation for a positive radial electric field while they slow down ion rotation for a negative electric field. 
For a given electric field amplitude, the ion angular frequency $|\varpi_i |$ is therefore larger in the case of a positive electric field. For this reason, positive fields appear more promising for applications where fast rotation is desirable, such as mass separation. 

The slow Brillouin mode~\cite{Davidson2001,Rax2015}, which describes the collisionless rigid-rotor equilibrium solution of a plasma rotating under a positive radial electric field, indicates that $\varpi/\Omega_i = \mathcal{O}(1)$ for sufficiently large radial electric field. For the Brillouin limit, which corresponds to the maximum electric field for which ions are still radially confined, one finds $\varpi_i=2\varpi_e = -\Omega_i/2$. Such angular frequencies are well beyond the rotation speed regimes studied by Chen~\cite{Chen1966}. To the extent that centrifugal instability stems from a difference in equilibrium azimuthal velocity, and that this difference grows rapidly as the rotation gets faster and approaches the Brillouin limit, fast rotation is expected to alter the characteristics of this instability.

In this paper, we study the stability of a magnetized plasma column rotating under a positive radial electric field with an eye on the fast rotation regime. In this study, fast rotation is defined as $|\varpi_i -\varpi_e |/\Omega_i\sim \mathcal{O}(1)$, that is to say when the slow Brillouin solution differs significantly from the massless limit. In Sec.~\ref{Sec:II}, the equilibrium rotation profiles are derived using a two-fluid formalism, and the differences between Simon-Hoh and centrifugal instabilities are highlighted. In the process, the Brillouin modes modified by collisions are recovered. In Sec.~\ref{Sec:III}, the dispersion relation for the centrifugal instability in cylindrical geometry is presented and discussed. In Sec.~\ref{Sec:IV}, theoretical findings are used to highlight experiments where this instability might be found. In Sec.~\ref{Sec:V}, the main findings are summarized.



\section{Equilibrium profiles}
\label{Sec:II}

Consider a radially bounded plasma in axisymetric geometry immersed in an axial magnetic field $\mathbf{B} = B_0\mathbf{\hat{z}}$. Let us assume the magnetic field constant across the plasma radius. Furthermore, the density and potential depend only on the radial coordinate $r$. Adopting a two-fluid formalism, the governing equations are the momentum equation for ions
\begin{multline}
m_i n_i\left(\frac{\partial \mathbf{v}_i}{\partial t} + \mathbf{v}_i\cdot\bm{\nabla}\mathbf{v}_i\right) =  e n_i\left(-\bm{\nabla}\phi +\mathbf{v}_i\times\mathbf{B}\right)\\-m_i n_i \nu_i \mathbf{v}_i,
\label{Eq:ion_momentum}
\end{multline}
the momentum equation for electrons
\begin{equation}
0 = -e n_e \left(-\bm{\nabla}\phi+\mathbf{v}_e\times\mathbf{B}\right) - k_bT_e\bm{\nabla}n_e - m_e n_e \nu_e \mathbf{v}_e,
\label{Eq:electron_momentum}
\end{equation}
and the continuity equation for each species
\begin{equation}
\frac{\partial n_j}{\partial t} +\bm{\nabla}\cdot(n_j\mathbf{v}_j) = 0.
\end{equation}
Here $j=i,e$ denotes either ions or electrons, and $\nu_{e}$ and $\nu_{i}$ are the electron-neutral and ion-neutral collision frequencies, respectively. Ions are assumed to be cold ($T_i = 0$), electron-neutral and ion-neutral collisions are supposed to dominate over Coulomb collisions, and electron inertia is neglected.

In the small azimuthal velocity limit, the convection term in the ion momentum equation can be neglected, and one recovers the classic formula
\begin{equation}
v_{i_{\theta}} = \frac{1}{1+\frac{\nu_{i}^2}{\Omega_i^2}}\frac{\phi^{\prime}}{B_0}~, 
\label{Eq:ion_SH}
\end{equation}
where a prime denotes $\partial/\partial r$, and $\Omega_i = eB_0/m_i$ is the ion cyclotron frequency. A similar expression is obtained for the electrons when neglecting the diamagnetic drift velocity resulting from the electron pressure term. In this limit, the classical Simon-Hoh~\cite{Simon1963,Hoh1963} instability arises when $\phi^{\prime}\cdot n^{\prime}<0$, granted that the collisionality is such that $\Omega_e/\nu_{e}\gg1$ and $\Omega_i/\nu_{i}\geq 1$. This instability stems from the larger azimuthal velocity of electrons, $v_{e_{\theta}}\sim\phi'/B_0$, as compared to the one of ions defined in Eq.~(\ref{Eq:ion_SH}).  

For large ion velocities, the convection term can no longer be neglected, and plugging the $\theta$ component into the $r$ component of Eq.~(\ref{Eq:ion_momentum}) gives the following equation for the azimuthal ion velocity:
\begin{widetext}
\begin{multline}
{v_{i_{\theta}}}^{2}+v_{i_{\theta}}r\Omega_i \left(1+\left(\frac{\nu_{i}}{\Omega_i}\right)^2\frac{1}{1+\frac{v_{i_{\theta}}}{r\Omega_i}+\frac{{v_{i_{\theta}}}^{\prime}}{\Omega_i}}-\left(\frac{\nu_{i}}{\Omega_i}\right)^2\frac{\left(\frac{v_{i_{\theta}}}{r\Omega_i}\right)^2+\frac{{v_{i_{\theta}}}^{\prime}}{\Omega_i}\left(1+\frac{{v_{i_{\theta}}}^{\prime}}{\Omega_i}\right)-\frac{v_{i_{\theta}}{v_{i_{\theta}}}^{\prime\prime}}{{\Omega_i}^2}}{\left(1+\frac{v_{i_{\theta}}}{r\Omega_i}+\frac{{v_{i_{\theta}}}^{\prime}}{\Omega_i}\right)^3}\right)
-r\Omega_i\frac{\phi^{\prime}}{B_0} = 0.
\label{Eq:ion_momentum_r}
\end{multline}
\end{widetext}
In Eq.~(\ref{Eq:ion_momentum_r}), the second and third terms inside the parentheses are the collision term and the $v_{r_i}{v_{r_i}}^{\prime}$ component of the convective derivative, respectively. Bringing these terms together, Eq.~(\ref{Eq:ion_momentum_r}) can be rewritten in a more compact form
\begin{widetext}
\begin{equation}
{v_{i_{\theta}}}^{2}+v_{i_{\theta}}r\Omega_i \left(1+\left(\frac{\nu_{i}}{\Omega_i}\right)^2\frac{1+\frac{{v_{i_{\theta}}}^{\prime}}{\Omega_i}+\frac{v_{i_{\theta}}}{r\Omega_i}\left(2+2\frac{{v_{i_{\theta}}}^{\prime}}{\Omega_i}+\frac{r{v_{i_{\theta}}}^{\prime\prime}}{\Omega_i}\right)}{\left(1+\frac{v_{i_{\theta}}}{r\Omega_i}+\frac{{v_{i_{\theta}}}^{\prime}}{\Omega_i}\right)^3}\right)\\
-r\Omega_i\frac{\phi^{\prime}}{B_0} = 0. 
\label{Eq:ion_momentum_r2}
\end{equation}
\end{widetext}
For low ion collisionality regimes ($\nu_i/\Omega_i \ll 1$), Eq.~({\ref{Eq:ion_momentum_r2}) reduces to a second order equation for $v_{i_{\theta}}$. Its solution are the slow and fast Brillouin modes~\cite{Davidson2001},
\begin{equation}
v_{i_{\theta}} = -r\frac{\Omega_i}{2}\left(1\mp\sqrt{1+4\frac{\phi^{\prime}}{r \Omega_i B_0}}\right). 
\label{Eq:Brillouin}
\end{equation}
Introducing $p = -4\phi^{\prime}/(r \Omega_i B_0)$, the slow mode rotation speed up is 
\begin{equation}
\iota_i = \frac{v_{i_{\theta}} }{\phi^{\prime}/B_0} = \frac{2}{p}(1-\sqrt{1-p}),
\end{equation}
and $\iota_i=2$ for the Brillouin limit $p=1$.

When collisionality is not negligible, Eq.~(\ref{Eq:ion_momentum_r2}) can be further simplified if assuming solid body rotation, that is to say
\begin{subequations}
\begin{align}
{v_{i_{\theta}}} & = \varpi_i r\\
{\varpi_i}^{\prime} & = {\varpi_i}^{\prime\prime} = 0.
\end{align}
\end{subequations}
For this simple rotation profile, Eq.~(\ref{Eq:ion_momentum_r2}) becomes
\begin{equation}
{\varpi_i}^{2}+\varpi_i\Omega_i \left(1+\left(\frac{\nu_{i}}{\Omega_i}\right)^2\frac{1+{\varpi_i}/\Omega_i}{\left(1+2\varpi_i/\Omega_i\right)^2}\right)\\
-\Omega_i\frac{\phi^{\prime}}{r B_0} = 0, 
\end{equation}
and an exact solution for $\Theta = \varpi_i/\Omega_i$ can be found in the form of the roots of the fourth order polynomial equation
\begin{multline}
4\Theta^4+8\Theta^3+\Theta^2\left[5+\left(\frac{\nu_i}{\Omega_i}\right)^2-4\frac{\phi^{\prime}}{r\Omega_i B_0}\right]\\
+\Theta\left[1+\left(\frac{\nu_i}{\Omega_i}\right)^2-4\frac{\phi^{\prime}}{r\Omega_i B_0}\right] - \frac{\phi^{\prime}}{r B_0 \Omega_i} = 0.
\end{multline}
Using the change of variable $\alpha = -(1+2 \Theta)\Omega_i/\nu_i$, one recovers the quartic equation~\cite{Rax2015}
\begin{equation}
\alpha^4+\left[1-\left(\frac{\Omega_i}{\nu_i}\right)^2-4\frac{\phi^{\prime}\Omega_i}{r{\nu_i}^2 B_0}\right]\alpha^2-\left(\frac{\Omega_i}{\nu_i}\right)^2 = 0. 
\end{equation}
The solution $\Theta = v_{i_{\theta}}/(r\Omega_i)$ is plotted in Figure~\ref{Fig:CollisionalModes} as a function of the radial electric field ($\phi^{\prime}<0$) for various $\nu_i/\Omega_i$ values. Although collisions slow down the angular frequency of the slow rotation mode, $|v_{i_{\theta}}|$ remains larger than $|\phi^{\prime}/B_0|$ as long as $\nu_i/\Omega_i \leq 0.3$. 

\begin{figure}
\includegraphics{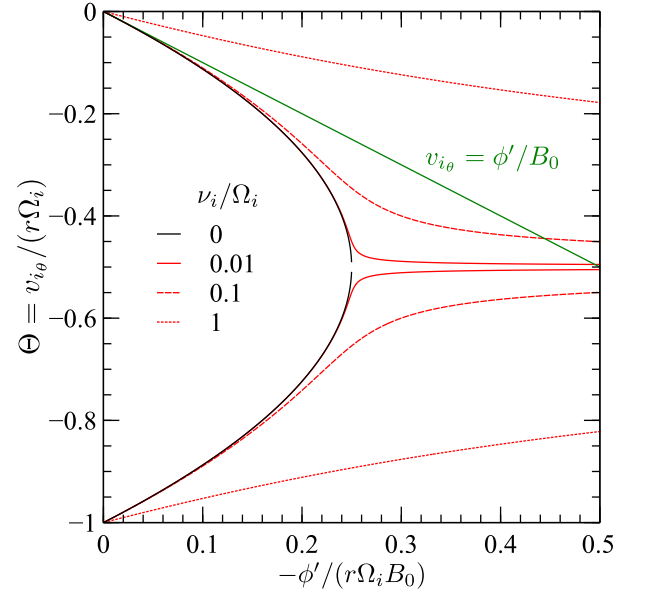}
\caption{Normalized angular frequencies $\Theta = v_{i_{\theta}}/(r\Omega_i)$ of the two rotation modes for the solid body rotation case as function of $p/4 = -\phi^{\prime}/(r\Omega_i B_0)$  for different collisionality regimes: $\nu_{i}/\Omega_i = 0$ (blue), and $\nu_{i}/\Omega_i = 0.01,0.1$ and $1$ (red). The massless limit ($p\rightarrow 0$) $v_{i_{\theta}} = \phi^{\prime}/B_0$ is plotted in green. }
\label{Fig:CollisionalModes}
\end{figure}

Looking now at electrons, one gets $|\phi^{\prime}/(r\Omega_eB_0)|\ll1$ since $m_i/m_e\geq 1836$. Centrifugal effects are therefore negligible in first approximation. As a matter of fact, for $4\phi^{\prime}/(r\Omega_i B_0) \sim -1$, \emph{i.~e.} the maximum value for ion confinement in a non collisional plasma, one gets $-\phi^{\prime}/(r\Omega_e B_0) \leq 1.5~10^{-4}$. Eq.~(\ref{Eq:electron_momentum}) then yields
\begin{align}
v_{e_\theta} & = \frac{1}{1+\left(\frac{\nu_e}{\Omega_e}\right)^2}\left[\frac{\phi^{\prime}}{B_0}-\frac{k_B T_e}{eB_0}\frac{{n_e}^{\prime}}{n_e}\right]\nonumber\\
 &\sim\frac{\phi^{\prime}}{B_0}-\frac{k_B T_e}{eB_0}\frac{{n_e}^{\prime}}{n_e}  = v_{E\times B}+v^{\star},
\label{Eq:electron_azimuthal_velocity}
\end{align}
where we have used the ordering $\nu_e/\Omega_e\ll1$. Here $v_{E\times B}$ and $v^{\star}$ are respectively the $E\times B$ drift velocity and the electron diamagnetic drift velocity. 

Neglecting in first approximation the effects of electron pressure and collisions, electrons and ions display different azimuthal velocities as a result of the larger effect of centrifugal forces on ions. Quantitatively, $v_{i_{\theta}}\leq v_{e_\theta}\leq 0$, and 
\begin{equation}
\frac{v_{i_{\theta}}-v_{e_{\theta}}}{r\Omega_i} = -\frac{1}{2}\left(1-\frac{p}{2}-\sqrt{1-p}\right).
\label{Eq:diff}
\end{equation}
In Eq.~(\ref{Eq:diff}),  $|v_{E\times B}| \gg |v^{\star}|$ has been assumed, and collisions have been neglected. For $p\rightarrow 0$, 
\begin{equation}
\frac{v_{i_{\theta}}-v_{e_{\theta}}}{r\Omega_i}\sim -\left(\frac{v_{e_{\theta}}}{r\Omega_i}\right)^2\ll1,
\end{equation}
which is the regime considered by Chen~\cite{Chen1966}. In contrast, for the Brillouin limit $p=1$, one gets 
\begin{equation}
\frac{v_{i_{\theta}}-v_{e_{\theta}}}{\Omega_i r} = \frac{v_{e_{\theta}}}{\Omega_i r}= -\frac{1}{4},
\end{equation}
so that the ion azimuthal velocity is twice as large as the electron azimuthal velocity. 

As opposed to the classical Simon-Hoh instability (see~Table.~\ref{Tab:tab1}), charged particles orbit here in the clockwise direction, and the amplitude of the ion azimuthal velocity is larger than the one of electrons. However, since both the rotation direction and the azimuthal velocity ordering are reversed, this configuration is also unstable for $n^{\prime}<0$. The difference in azimuthal velocities causes a space charge separation between ion and electron density perturbations in the azimuthal direction, therefore producing a positive azimuthal electric field $E_{\theta}$. For $n^{\prime}<0$, the resulting $\mathbf{E}_{\theta}\times\mathbf{B}$ enhances the density perturbation.

\begin{table}[t]
\begin{center}
\begin{tabular}{c | c | c}
 & \shortstack{Classical Simon-Hoh \\ instability}  & \shortstack{Fast rotation \\ centrifugal instability} \\ 
\hline
$\phi^{\prime}$ & Positive & Negative\\
Rotation direction & Anti-clockwise & Clockwise\\
Fastest species & Electrons & Ions\\
$n^{\prime}$ for instability & Positive & Positive\\
\end{tabular}
\end{center}
\caption{Comparison of the main features of the classical Simon-Hoh instability~\cite{Simon1963,Hoh1963} and of the fast rotation centrifugal instability described in this study.}
\label{Tab:tab1}
\end{table}
 
 \section{Perturbation in cylindrical geometry}
 \label{Sec:III}

The dispersion relation for the centrifugal instability in the fast rotation regime is derived, in cylindrical geometry, in Appendix~\ref{Sec:AppDerivation}.  In the following, $\upsilon_0$ indicates the equilibrium value of variable $\upsilon$, while $\tilde{\upsilon}$ denotes the perturbative part of $\upsilon$. Variables used throughout the derivation are given in Table~\ref{Tab:tab3}. Under the hypotheses that (\emph{i}) ions are cold ($T_i\sim 0$), so that finite Larmor radius stabilization effects~\cite{Hoh1963a,Chen1966} are not considered, (\emph{ii}) collisions are negligible ($\nu_{e} = \nu_{i}=0$), (\emph{iii}) equilibrium rotation profile is not sheared (${\varpi_0}'=0$) and (\emph{iv}) $k_z \sim 0$, the perturbed densities write:
\begin{subequations}
\begin{align}
\frac{\tilde{n}_e}{n_0} & = \chi\frac{\omega^{\star}}{\omega-\omega_E} \label{Eq:electron_density_perturbationM}\\
 \frac{\tilde{n}_i}{n_0} & = \chi \frac{{c_s}^2}{r^2(\omega-\omega_B)}\frac{m^2(\omega-\omega_B) -m\sqrt{1-p}\Omega_i\lambda}{(\omega-\omega_B)^2-(1-p){\Omega_i}^2}.
 \end{align}
\end{subequations}
Here ${c_s} = \sqrt{k_B T_e/m_i}$ is the ion sound speed, $\chi = e\tilde{\phi}/(k_B T_e)$ is the normalized perturbed potential, $\lambda = r |{n_0}'/n_0|$ is the dimensionless density gradient scale-length and $p = -4{\phi_0}^{\prime}/(r \Omega_i B_0)$ is the Brillouin parameter. Invoking quasi-neutrality ($\tilde{n}_e=\tilde{n}_i$), one gets a cubic equation $\bar{\omega}^3+\alpha_2 \bar{\omega}^2 + \alpha_1 \bar{\omega} + \alpha_0 = 0$ for the normalized complex frequency $\bar{\omega} = \omega/\Omega_i = \bar{\omega}_r+i\bar{\gamma}$, where the coefficients are real and equal to
\begin{subequations}
\label{Eq:coefficients}
\begin{align}
\alpha_2 &= m(3\lambda\zeta-2)/(2\lambda),\\
\alpha_1 &= \zeta \left[4\lambda(1-\zeta)+m^2(3\lambda\zeta+\zeta-4)\right]/(4\lambda),\\
\alpha_0 &= m\zeta^2\left[2\lambda(1-\zeta)+m^2(\zeta\lambda+\zeta-2)\right]/(8\lambda),
\end{align}
\label{Eq:cubic}
\end{subequations}
with $\bar{\omega}^{\star} = \omega^{\star}/\Omega_i$ and $\zeta = 1-\sqrt{1-p}$. 
Stability is determined by the positiveness of the discriminant 
\begin{multline}
\Delta = \frac{\zeta^2}{16\lambda^4}\left[64(\zeta-1)^3\zeta\lambda^4+m^6\zeta^2(1-\zeta^2\lambda)\right.\\\left.+2m^4(\zeta-1)\lambda
\left(8+\zeta\left[3\zeta\lambda(2\zeta-3)-4\right]\right)\right.\\\left.-m^2(\zeta-1)^2\lambda^2\left(3\zeta\lambda\left[\zeta(9\lambda+16)-24\right]-16\right)\right]
\label{Eq:Discriminant}
\end{multline}
of this cubic equation. Stable and unstable regions obtained based on this criteria are plotted in Fig.~\ref{Fig:Discr} as a function of a function of the Brillouin parameter $p$ and the dimensionless density gradient scale-length $\lambda$. 

 \begin{table*}[htb]
\begin{center}
\begin{tabular}{c | c | c}
Variable & Notation & Definition \\ 
\hline
Ion gyro-frequency & $\Omega_i$ & $eB_0/m_i$\\
Ion angular velocity & $\varpi_i$ & $v_{i_{\theta}}/r$\\
Ion sound speed & $c_s$ & $ \sqrt{k_B T_e/m_i}$\\
Electron diamagnetic angular frequency & $\omega^{\star}$ & $ -(m/r)({n_0}'/n_0)(k_B T_e)(e B_0)$\\
Brillouin angular frequency & $\omega_B$ & $m\varpi_i$\\
$\mathbf{E}\times\mathbf{B}$ angular frequency & $\omega_E$ & $(m/r)({\phi_0}'/B_0)$\\
Perturbation frequency & $\omega_r$ & $\Re(\omega)$\\
Perturbation growth-rate & $\gamma$ & $\Im(\omega)$\\
Perturbation azimuthal mode number & $m$ & $k_{\theta}r$\\
Brillouin parameter & $p$ & $-4{\phi_0}^{\prime}/(r \Omega_i B_0)$\\
Normalized ion angular velocity & $\zeta$ & $1-\sqrt{1-p}$\\
Normalized perturbed potential & $\chi$ & $e\tilde{\phi}/(k_B T_e)$\\
Normalized density gradient scale-length & $\lambda$ & $r |{n_0}'/n_0|$
\end{tabular}
\end{center}
\caption{Main dimensional and dimensionless variables used.}
\label{Tab:tab3}
\end{table*}

\begin{figure}
\subfigure[~$m=1$]{\hspace{-0.3cm}\includegraphics{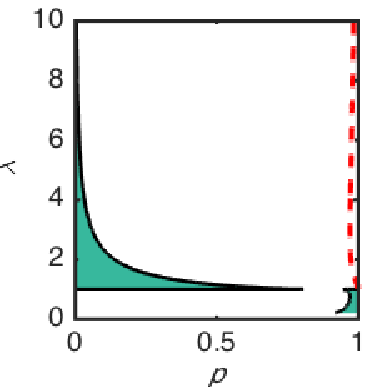}\label{Fig:Stability_Diagram_m_1}}\subfigure[~$m=2$]{\hspace{-0.3cm}\includegraphics{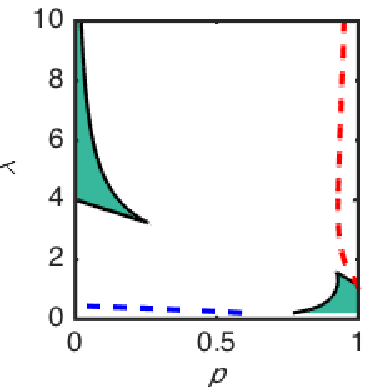}\label{Fig:Stability_Diagram_m_2}}\\
\subfigure[~$m=3$]{\hspace{-0.3cm}\includegraphics{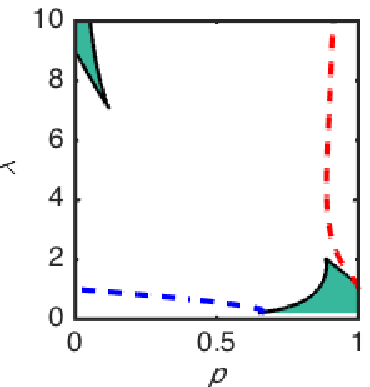}\label{Fig:Stability_Diagram_m_3}}\subfigure[~$m=5$]{\hspace{-0.3cm}\includegraphics{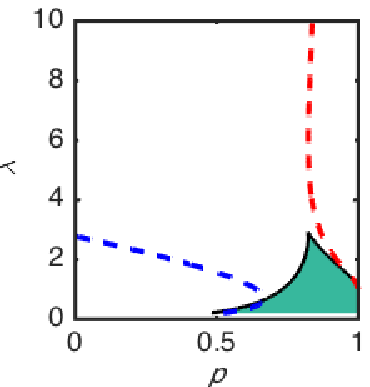}\label{Fig:Stability_Diagram_m_5}}\\
\subfigure[~$m=10$]{\hspace{-0.3cm}\includegraphics{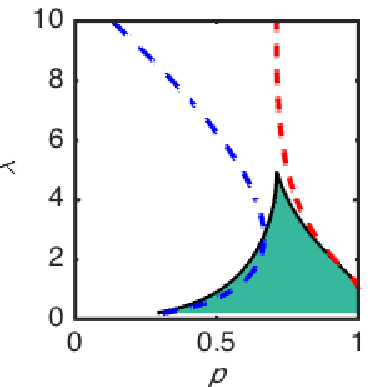}\label{Fig:Stability_Diagram_m_10}}\subfigure[~$m=30$]{\hspace{-0.3cm}\includegraphics{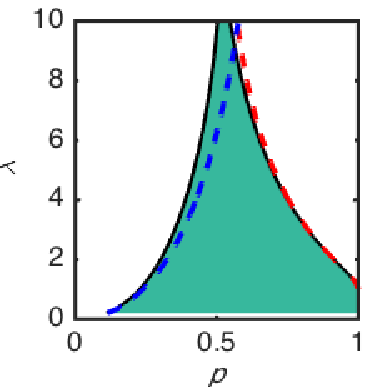}\label{Fig:Stability_Diagram_m_30}}
\caption{Stability diagram as a function of the Brillouin parameter $p = -4{\phi_0}^{\prime}/(r \Omega_i B_0)$ and the dimensionless density gradient scale-length $\lambda = r |{n_0}'/n_0|$ (see Table~\ref{Tab:tab3}). The green shaded region is the stable region. The dashed blue curve depicts the small $p$, $m\gg\lambda$ limit given by Eq.~(\ref{Eq:small_p}). The dashed red curve depicts the $m(\bar{\omega}-\bar{\omega}_B)\gg\lambda$ limit given by Eq.~(\ref{Eq:High_m_regime}). }
\label{Fig:Discr}
\end{figure}

\subsection{Asymptotic regimes}

Various asymptotic regimes can be identified for Eq.~(\ref{Eq:cubic}). For $p=0$, $\omega_B = \omega_E =  0$, quasi-neutrality reduces to
\begin{equation}
\omega^{\star} = \frac{{c_s}^2}{r^2}\frac{m^2\omega-m\lambda\Omega_i}{\omega^2-{\Omega_i}^2},
\end{equation}
which corresponds to the stable anti-drift mode~\cite{Fridman1964}
\begin{equation}
\omega = \frac{m^2{c_s}^2}{r^2\omega^{\star}}.
\end{equation}
Therefore no instability exists for $p=0$. Expanding now the solutions of Eq.~(\ref{Eq:cubic}) to the first order in $p$, the two complex conjugate roots write 
\begin{equation}
\bar{\omega} = \left[-(m^2-\lambda) \pm i\sqrt{(m^2-\lambda)\lambda}\right]\frac{p}{4m} +\mathcal{O}(p^2),
\end{equation}
which is unstable for $m^2>\lambda$ granted that $p>0$, as seen in Fig.~\ref{Fig:Discr}. The growth rate in the slow rotation limit ($p\ll 1$) 
\begin{equation}
\bar{\gamma} = \frac{\sqrt{(m^2-\lambda)\lambda}}{4m}p \underset{m^2\gg\lambda}{\rightarrow} \frac{\sqrt{\lambda}}{4}p
\label{Eq:slow_rotation_growth_rate}
\end{equation}
is therefore proportional to $p$ and independent of the mode number $m$ for $m^2\gg \lambda$. In contrast, the real frequency is proportional to both $m$ and $p$, and writes 
\begin{equation}
\omega_r = \omega_B + \frac{{\Omega_i}^2}{4{k_{\theta}}^2{c_s}^2}\omega^{\star}p.
\end{equation}

Although first order expansion suggests an instability for $p>0$, Taylor expanding the discriminant in Eq.~(\ref{Eq:Discriminant}) for small $p$,
\begin{equation}
\Delta = \mathcal{D}_2 p^2 +\mathcal{D}_3 p^3 +\mathcal{D}_4 p^4 +\mathcal{O}(p^5),
\end{equation}
with
\begin{subequations}
\begin{align}
\mathcal{D}_2  = & -m^2(m^2-\lambda)/(4\lambda^3)\\
\mathcal{D}_3  = & \left[3 m^4-8\lambda^3+m^2\lambda(9\lambda-2)\right]/(16\lambda^3)\\
\mathcal{D}_4  = & \left(m^6+96 \lambda^4+2m^4\lambda(9\lambda+2)\right.\nonumber\\
 & \left.+m^2\lambda^2\left[4-3\lambda(9\lambda+52)\right]\right)/(256\lambda^4),
 \end{align}
\end{subequations} 
reveals that the instability ceases for $p>p_s$, with
\begin{equation}
p_s  \underset{m\gg\lambda}{\rightarrow} \frac{8\sqrt{\lambda}}{m}\left(1-3\frac{\sqrt{\lambda}}{m}\right).
\label{Eq:small_p}
\end{equation}
The solution of Eq.~(\ref{Eq:small_p}) is plotted in dashed blue in Fig.~\ref{Fig:Discr}. For $m=10$ and $\lambda=1$, this gives $p_s\sim0.56$, which is close to $p=0.52$ for which transition is observed in Fig.~\ref{Fig:Stability_Diagram_m_10}. 



On the other hand, for fast rotation at the Brillouin limit, \emph{i.~e.} $p=1$, quasi-neutrality writes 
\begin{equation}
\frac{\omega^{\star}}{\omega-\omega_E} = \frac{m^2{c_s}^2}{(\omega-\omega_B)^2},
\end{equation}
and
\begin{equation}
\omega_{1,2}  = \omega_B+\frac{{c_s}^2{k_{\theta}}^2}{2\omega^{\star}}\pm\sqrt{\frac{1}{4}\left(\frac{{c_s}^2{k_{\theta}}^2}{\omega^{\star}}\right)^2+\frac{{c_s}^2{k_{\theta}}^2}{\omega^{\star}}(\omega_B-\omega_E)},
\end{equation}
or, equivalently,
\begin{equation}
\bar{\omega}_{1,2} = \frac{m}{2\lambda}(1-\lambda)\left[1\pm \sqrt{\frac{1}{1-\lambda}}\right].
\end{equation}
Here we have used the relation $\bar{\omega}_B = 2\bar{\omega}_E = -m/2$ which is valid for $p=1$. In this limit, an instability is therefore found for any mode number $m$ as long as $\lambda>1$, or
\begin{equation}
\omega_B-\omega_E<-\frac{{c_s}^2{k_{\theta}}^2}{4\omega^{\star}},
\end{equation} 
and the growth rate is
\begin{equation}
\bar{\gamma} = \frac{m}{2}\sqrt{\frac{\lambda-1}{\lambda^2}} \underset{\lambda \gg1}{\rightarrow}  \frac{m}{2\sqrt{\lambda}}.
\label{Eq:growth_rate_Brillouin}
\end{equation}
For fast rotation but $p\neq1$, a more general result can be obtained for $\lambda\ll m|(\bar{\omega}-\bar{\omega}_B)|$. In this limit, the complex frequency is solution of the equation  $\bar{\omega}^2+\beta_1 \bar{\omega} + \beta_0 = 0$ (see Appendix~\ref{Sec:AppDerivation}), with
\begin{subequations}
\begin{align}
\beta_1 &= m(\zeta-\lambda^{-1}),\\
\beta_0 &=  m^2\zeta(\zeta\lambda+\zeta-2)/(4\lambda)-(\zeta-1)^2.
\end{align}
\label{Eq:High_m_regime}
\end{subequations}
The instability threshold obtained from this asymptotic regime is plotted in dashed red in Fig.~\ref{Fig:Discr}, and agrees well, as expected, with the general solution for large $p$ and $m\gg\lambda$ (\emph{i.~e.} $k_{\theta}\gg k_{ne}$). In this limit, the growth rate is 
\begin{equation}
\bar{\gamma} = \frac{\sqrt{m^2\left[(1-\sqrt{s})^2\lambda-1\right]-4s\lambda^2}}{2\lambda} \underset{\lambda\gg1}{\rightarrow} \frac{m}{2\sqrt{\lambda}}(1-\sqrt{s}),
\label{Eq:High_m_regime_growth_rate}
\end{equation}
where $s = 1-p$, which is consistent with Eq.~(\ref{Eq:growth_rate_Brillouin}). In contrast with the slow rotation limit given by Eq.~(\ref{Eq:slow_rotation_growth_rate}), the growth rate for fast rotation is proportional to the azimuthal mode number $m$.

 \subsection{General solution}

The general stability picture obtained from the full solution is depicted in Fig.~\ref{Fig:Discr}. For low azimuthal mode number $m$, stability is only found for large enough $p$ for $\lambda<1$, and small enough $p$ for large $\lambda$ (see Figs.~\ref{Fig:Stability_Diagram_m_1}, \ref{Fig:Stability_Diagram_m_2} and \ref{Fig:Stability_Diagram_m_3}). For larger $m$, a stable region forms for intermediate $p$ (see Figs.~\ref{Fig:Stability_Diagram_m_5}, \ref{Fig:Stability_Diagram_m_10} and \ref{Fig:Stability_Diagram_m_30}), between the slow and fast asymptotic regime given by Eq.~(\ref{Eq:small_p}) and Eq.~(\ref{Eq:High_m_regime}), respectively.
 
Figure~\ref{Fig:Growth_rate} shows the normalized real frequency $\bar{\omega}_r$ and growth-rate $\bar{\gamma}$ as obtained by solving numerically the cubic equation defined by Eq.~(\ref{Eq:cubic}). Although instability is found over a large fraction of the $(p,\lambda)$ parameter space studied, the growth-rate $\gamma$ of this instability varies greatly over this region. Overall the growth rate is found to grow with $p$. This is particularly true for large mode numbers (see, \emph{e.~g.}, fig.~\ref{Fig:Growth_rate_m_10}) where a strong increase of the growth-rate is observed as $p$ approaches $1$. This feature is consistent with the $1-\sqrt{1-p}$ scaling found fo the fast rotation asymptotic regime given in Eq.~(\ref{Eq:High_m_regime_growth_rate}), and is particularly visible in Fig.~\ref{Fig:Growth_Rate} for $m\geq 3$ and $p\geq0.9$. Intuitively, the increase of $\gamma$ with $p$ is consistent with the greater difference between electrons and ions equilbrium azimuthal velocities $v_{e_\theta}$ and $v_{i_\theta}$, as shown by Eq.~(\ref{Eq:diff}). In addition, Fig.~\ref{Fig:Growth_Rate} confirms that the growth rate at low $p$ is independent of $m$, whereas it scales with $m$ at large $p$ to reach $\bar{\gamma}= 0.25~m$ for $p=1$. These results are consistent with the slow and fast rotation asymptotic solutions given respectively  by Eq.~(\ref{Eq:slow_rotation_growth_rate}) and  Eq.~(\ref{Eq:growth_rate_Brillouin}).


 \begin{figure}
\subfigure{\hspace{-0.15cm}\includegraphics[]{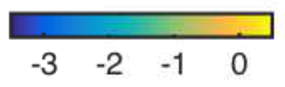}}
\addtocounter{subfigure}{-1}
\subfigure[~$\log_{10} \bar{\omega}_r, m=1$]{ \includegraphics[]{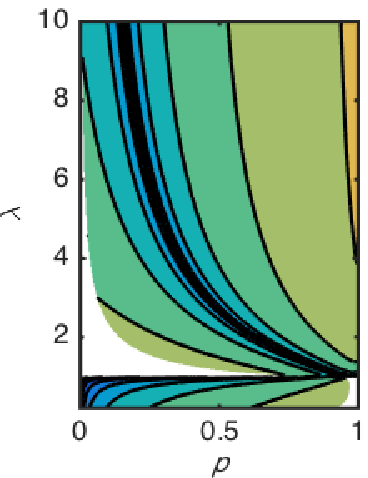}\label{Fig:Freq_m_1}}\subfigure[~$\log_{10} \bar{\gamma}, m=1$]{ \includegraphics[]{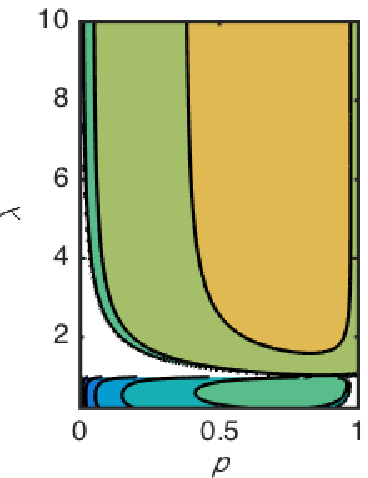}\label{Fig:Growth_rate_m_1}}\\
\subfigure[~$\log_{10} \bar{\omega}_r, m=3$]{ \includegraphics[]{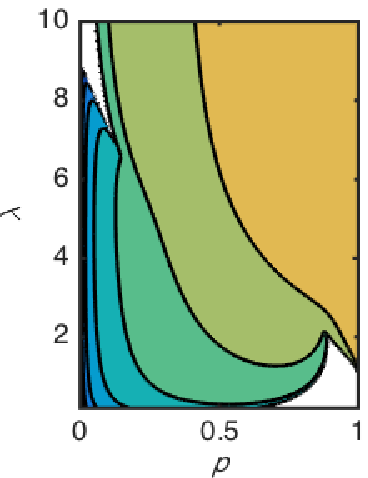}\label{Fig:Freq_m_3}}\subfigure[~$\log_{10} \bar{\gamma}, m=3$]{ \includegraphics[]{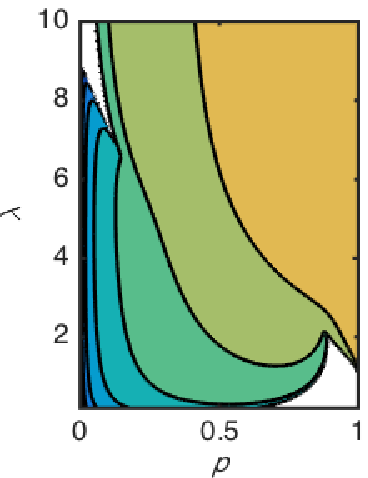}\label{Fig:Growth_rate_m_3}}\\
\subfigure[~$\log_{10} \bar{\omega}_r, m=10$]{ \includegraphics[]{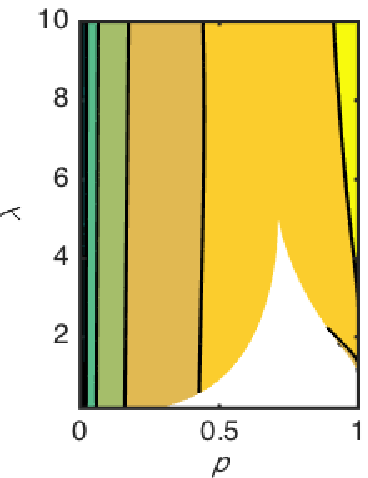}\label{Fig:Freq_m_10}}\subfigure[~$\log_{10} \bar{\gamma}, m=10$]{ \includegraphics[]{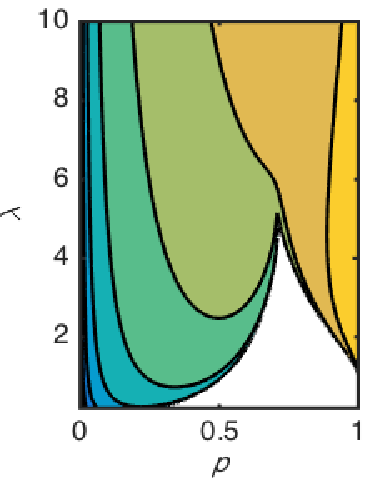}\label{Fig:Growth_rate_m_10}}
\caption{Normalized frequency $\bar{\omega} = \omega/\Omega_i$ [\subref{Fig:Freq_m_1}, \subref{Fig:Freq_m_3} and \subref{Fig:Freq_m_10}] and growth-rate $\bar{\gamma} = \gamma/\Omega_i$ [\subref{Fig:Growth_rate_m_1}, \subref{Fig:Growth_rate_m_3} and \subref{Fig:Growth_rate_m_10}] as a function of the Brillouin parameter $p = -4{\phi_0}^{\prime}/(r \Omega_i B_0)$ and the dimensionless density gradient scale-length $\lambda = r |{n_0}'/n_0|$ (see Table~\ref{Tab:tab3}) for three mode numbers. Only the domain where $\bar{\gamma}>0$ is plotted. }
\label{Fig:Growth_rate}
\end{figure}

\begin{figure}
\includegraphics{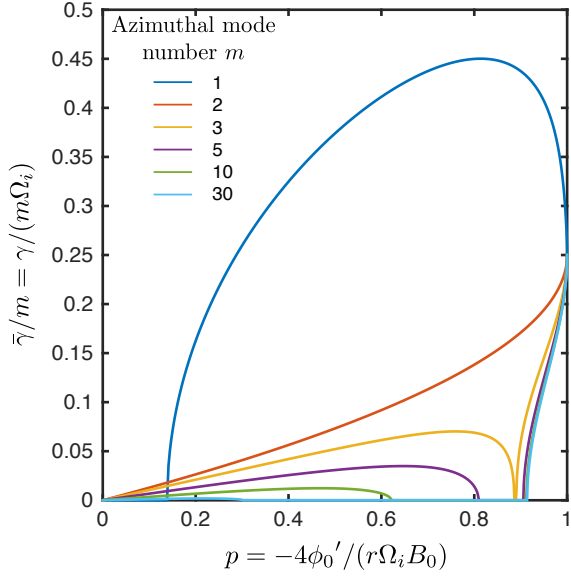}
\caption{Normalized growth rate $\bar{\gamma}/m = \gamma/(m\Omega_i)$ as a function of the Brillouin parameter $p = -4{\phi_0}^{\prime}/(r \Omega_i B_0)$ for various mode numbers.  The normalized density gradient used here is $\lambda = 2$. }
\label{Fig:Growth_Rate}
\end{figure}

For completeness, the growth rate $\gamma$ is plotted as a function of the mode number $m = k_{\theta} r$ in Fig.~\ref{Fig:Dispersion}. For a weak density gradient, for example $\lambda = 0.5$ as depicted in Fig.~\ref{Fig:Dispersion_Lambda_0_5}, the growth rate is small and decreases rapidly with $m$. The threshold mode number past which a decrease of $\gamma$ is observed decreases with $p$.  For stronger density gradient, for example $\lambda = 3$ shown in Fig.~\ref{Fig:Dispersion_Lambda_3}, the growth rate for low $m$ is significantly larger. Furthermore, the difference between fast and slow rotation regimes is clearly visible in Fig.~\ref{Fig:Dispersion}. Two different behaviours are observed depending on the value of $p$ compared to the instability threshold obtained in Eq.~(\ref{Eq:High_m_regime}) for $m(\bar{\omega}-\bar{\omega}_B)\gg\lambda$. As shown in dashed red in Fig.~\ref{Fig:Discr}, this threshold is about $0.82$ for $\lambda = 3$. For $p$ smaller than $0.82$, $\gamma$ is found to decrease with $m$, and a cut-off is found at large $m$. On the other hand, for $p$ larger than $0.82$, $\gamma$ grows linearly with $m$, and the slope grows with $p$. This dependency is once again consistent with the limit given by Eq~(\ref{Eq:High_m_regime_growth_rate}). 

\begin{figure}
\subfigure[~$\lambda=0.5$]{\includegraphics{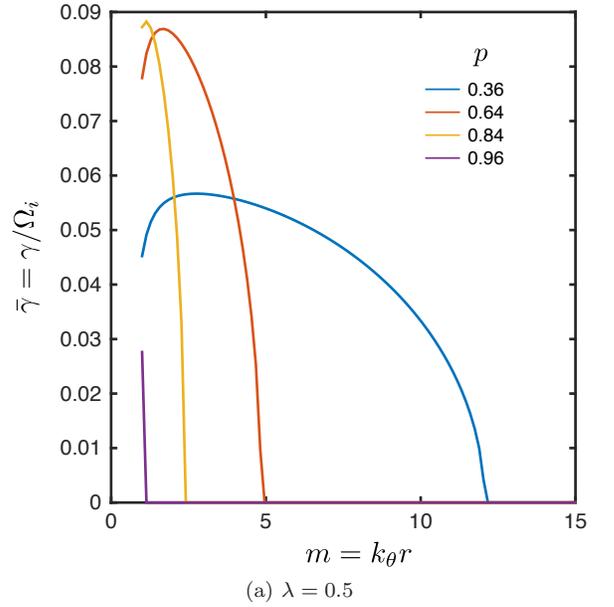}\label{Fig:Dispersion_Lambda_0_5}}
\subfigure[~$\lambda=3$]{\includegraphics{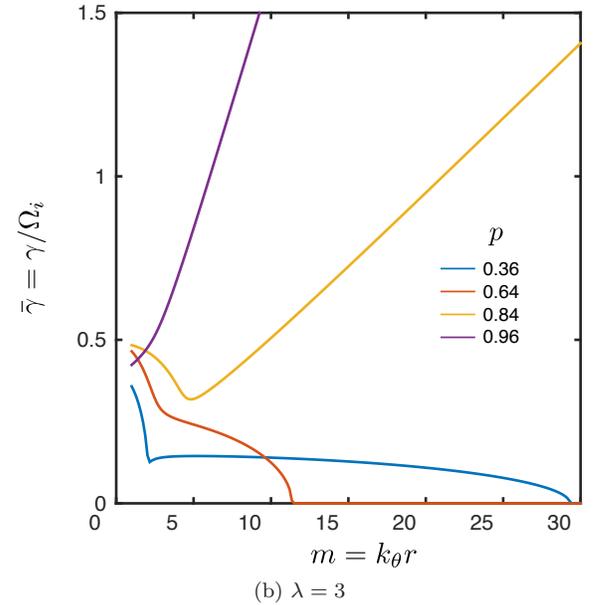}\label{Fig:Dispersion_Lambda_3}}
\caption{Growth rate for as a function of $m$ for a dimensionless density gradient scale-length $\lambda = r|{n_0}'/n0| = 0.5$ [\ref{Fig:Dispersion_Lambda_0_5}]  and $\lambda = 3$ [\ref{Fig:Dispersion_Lambda_3}] and different values of the Brillouin parameter $p = -4{\phi_0}^{\prime}/(r \Omega_i B_0)$. Differences between the slow (blue and red) and fast rotation (yellow and purple) regimes is clearly visible through the evolution at large $m$. }
\label{Fig:Dispersion}
\end{figure}

Although $\gamma$ appears to grow indefinitely with $m$ in Fig.~\ref{Fig:Dispersion_Lambda_3}, accounting for electron inertia leads to a cut-off of these modes at high $m$. As a matter of fact, for large $m$ the frequency is sufficiently large to invalidate the zero electron inertia assumption used up until now. In this case, Eq.~(\ref{Eq:electron_density_perturbationM}) has to be replaced by 
\begin{equation}
\frac{\tilde{n}_e}{n_0}  = \chi\frac{\omega^{\star}+{k_{\theta}}^2{\rho_e}^2(\omega-\omega_E)}{\omega-\omega_E}
\end{equation}
with $\rho_e = m_e {c_s}^2/(m_i{\Omega_i}^2)$ the electron Larmor radius. Assuming quasi-neutrality, the complex frequency $\omega$ is then solution of a fourth-order equation with real coefficients. Solving this equation shows that electron inertia results in a cut-off of the instability at high $m$ values, as illustrated in Fig.~\ref{Fig:Dispersion_Lambda_3_With_Inertia}. Nevertheless, there exists a range of azimuthal mode numbers for which the growth rate can be, in the limit of fast rotation, as large as a few ion cyclotron frequency. This is expected to be particularly true for heavy ions since electron inertia effects will set in for comparatively larger mode number.

\begin{figure}
\includegraphics{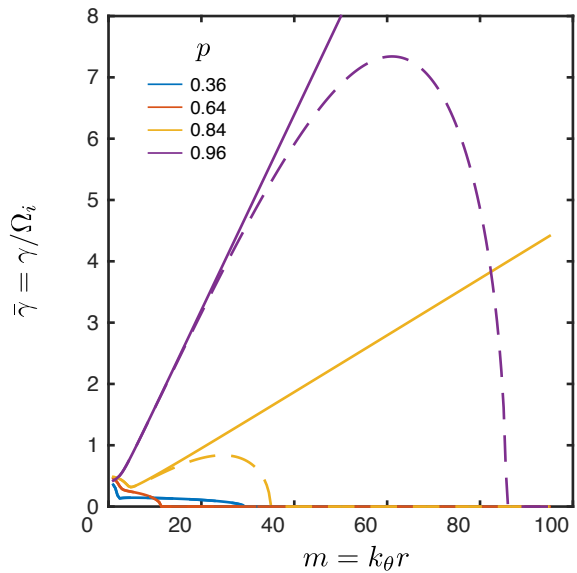}
\caption{Growth rate for as a function of $m$ for a dimensionless density gradient scale-length $\lambda = r|{n_0}'/n0| = 3$ and various values of the Brillouin parameter $p = -4{\phi_0}^{\prime}/(r \Omega_i B_0)$. Dashed line represent the solution accounting for electron inertia, while solid lines are from the cubic equation defined by Eq.~(\ref{Eq:cubic}). $m_i/m_e = 1836$. }
\label{Fig:Dispersion_Lambda_3_With_Inertia}
\end{figure}

In light of these results, it appears that the fastest growing instability is found for $p$ larger than the threshold value depicted in dashed red in Fig.~\ref{Fig:Discr}. In this regime, the growth-rate is shown to be as large as a few ion cyclotron frequency and to grow with both $p$ and $m$ until $\omega_r$ is large enough for electron inertia to become important. 

\section{Configurations suitable for this instability}
\label{Sec:IV}


Although a full literature review is beyond the scope of this study, it is informative to highlight which operating conditions are likely to favor the onset of this instability. A brief survey of the literature relevant to plasma mass separation suggests the fast centrifugal instability might develop in these devices.

Table~\ref{Tab:tab2} summarizes the operating parameters for three specific experiments. The first experiment~\cite{Shinohara2007} is a low pressure ($0.16$~mTorr), large helicon device operating in pure argon or pure xenon. The second experiment~\cite{Gueroult2016a} is a smaller helicon device operating mainly in argon (with a small fraction of krypton) at slightly higher pressure ($\sim5$~mTorr). The third one~\cite{Tsushima1986} is an ECR plasma operating in argon at $5~10^{-2}$~mTorr. All three devices are linear machines biased by means of concentric electrodes positioned at one or both ends of the device. Note however that the conditions listed in Table~\ref{Tab:tab2} are not necessarily found at the same spatial location, for example maximum density and potential gradient are not necessarily collocated.

\begin{table*}[htbp]
\begin{center}
\begin{tabular}{c | c | c | c}
 & Large Helicon device~\cite{Shinohara2007} & PMFX~\cite{Gueroult2016a}  & ECR plasma~\cite{Tsushima1986}\\ 
\hline
$\nu_i/\Omega_i$ & $3~10^{-4}$  & $10^{-2}$ & $10^{-4}$\\
$\max\left|{n_0}^{\prime}/n_0\right|$~[m$^{-1}$] & $60$ & $60$ & $60$\\
$\max|{\phi_0}^{\prime}|$~[V.m$^{-1}$] & $1500$ & $500$ & $300$\\
$B_0$~[kG]  & $1.2$  & $\leq 1$ & $1-3$\\ 
$m_i$~[amu] & $40-131$ & $40$ & $40$\\
$T_e$~[eV] & $3$ & $3$ & $5$\\
Typical bias applied [V] & up to $250$ & $20$ & $20$ \\
Discharge radius [cm] & $20$ & $10$ & $10$
\end{tabular}
\end{center}
\caption{Relevant parameters for three experiments of interest.}
\label{Tab:tab2}
\end{table*}

For all three experiments, the ion-neutral collision frequency is small compared to the ion gyro-frequency, and the ion temperature is negligible compared to the electron temperature. Consequently, the model derived in this study is expected to be valid, and centrifugal forces are expected to lead to different azimuthal velocities for electron and ions (see Eq.~(\ref{Eq:diff})), granted $-{\phi_0}^{\prime}/(rB_0\Omega_i) = p/4$ is non negligible. 

On the basis of the equilibrium density and potential profiles, the equivalent $(p,\lambda)$ parameters are computed for a set of representative biases for each experiment. The corresponding results are plotted in Fig.~\ref{Fig:Expe_data}, along with the instability threshold derived for the $m\gg\lambda$ regime and given by Eq.~(\ref{Eq:High_m_regime}). One immediately notices that some local operating conditions in the $(p,\lambda)$ domain exceed the threshold for instability. This is particularly true for the two biased helicon experiments (plus signs and open triangle symbols). Even for the ECR experiment (open circle symbols), operating conditions are locally found to approach the instability threshold. 

\begin{figure}
\includegraphics{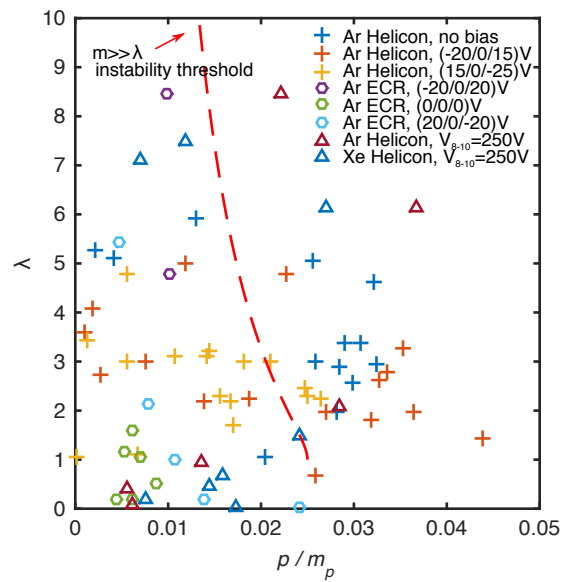}
\caption{Inferred $(p,\lambda)$ operating conditions for three biasing experiments: large helicon~\cite{Shinohara2007}(open triangles), smaller helicon in gas mixture~\cite{Gueroult2016a} (plus signs) and ECR~\cite{Tsushima1986} (open circles). The various colors depict different biasing conditions. The dashed red line represents the instability threshold derived in the $m\gg\lambda$ regime as given by Eq.~(\ref{Eq:High_m_regime}). The largest growth rates are found to the right of this line. Abscissa is normalized by the ion atomic mass.}
\label{Fig:Expe_data}
\end{figure}

Although the proximity of the operating conditions to the instability threshold in these three experiment might appear surprising at first, fluctuations have actually been reported in the ECR experiment~\cite{Tsushima1986,Tsushima1991}, and also in unbiased helicon discharges~\cite{Light2001}.  Interestingly, the amplitude of the fluctuations in the ECR experiment increases with the strength of the positive radial electric field imposed at the electrodes~\cite{Tsushima1991}, which is consistent with the fact that the negative potential gradient case (blue open circles) appear, on average, closer to the instability threshold than both the zero and negative gradient cases (green and purple open circles, respectively) in Fig.~\ref{Fig:Expe_data}.

Additionally, various explanations can be given to justify the fact that centrifugal instabilities might not be observed experimentally even if local operating conditions in the $(p,\lambda)$ space exceed the instability threshold derived in this work. First, the derivation proposed in this paper relies on local approximation, whereas ions will have a finite radial excursion in the experiments discussed in this section. For this reason, a refined instability threshold taking into consideration these non-local effects would have to be derived to properly assert the stability of these configurations. Second, as stated earlier, the experimental $(p,\lambda)$ operating conditions used in Fig.~\ref{Fig:Expe_data} are not necessary collocated, so that fast rotation might be found in weak density gradient region and vice-versa. Finally, the rotation in these devices display shear, that is to say $\partial \varpi/\partial r\neq 0$. Sheared rotation can lead to a different kind of instability, namely Kelvin-Helmholtz instabilty~\cite{Kent1969,Jassby1972,Light2001}, which could then compete with and possibly mask the fast rotation instability discussed in this paper.

\section{Summary and Conclusions}
\label{Sec:V}

In this paper, the centrifugal instability resulting from a difference between the azimuthal drift velocity of ions and electrons is considered in the limit of fast rotation. In this study, fast rotation means that the difference between ion and electron equilibrium angular velocity is not negligible compared to the electron equilibrium angular velocity.

By deriving the appropriate dispersion relation in cylindrical geometry, the stability of a collisionless solid body rotating plasma with cold ions is studied as a function of the density gradient length ${k_{ne}}^{-1}$ and the Brillouin parameter $p$, which is defined as four times the ratio between the $\mathbf{E} \times \mathbf{B}$ angular drift velocity and the ion cyclotron frequency $\Omega_i$. The growth rate for this instability is found to be maximum when approaching the Brillouin limit $p=1$, which corresponds to the maximum difference between ion and electron angular frequency. For these conditions and $k_{\theta}\gg k_{ne}$, the growth rate $\gamma$ is proportional to the azimuthal mode number $m$ and to $1-\sqrt{1-p}$, and $\gamma/(m\Omega_i) = \mathcal{O}(1)$ for $p=1$. For large wavenumbers, electron inertia is shown to lead to a cut-off. However, growth rate of a few ion cyclotron frequency are observed before electron inertia sets in. 

A brief survey of the literature, in particular of the typical operating conditions in experiments relevant to plasma mass separation, reveals that this instability is likely to develop in these devices. However, detailed instability characterization will require extending the model derived here to account for non-local effects. In addition, to the extent that rotating plasmas often features shear, Kelvin-Helmholtz instability could compete in these devices with the fast centrifugal instability studied here.

Since centrifugal effects and Brillouin modes depend strongly on particle mass, an interesting extension of this work will consist in asserting whether this instability mechanism could be used to produce differential transport properties in a multi-ion species plasma.

\section*{Acknowledgements}
This work was supported, in part, by US DOE Contract No. DE-SC0016072.

The authors would like to thank A. I. Smolyakov, Y. Raitses, G. J. M. Hagelaar and I. E. Ochs for constructive discussions. \\

\appendix

\section{Dispersion relation for the fast centrifugal instability}
\label{Sec:AppDerivation}

Introducing
\begin{subequations}
\begin{equation}
\omega_B = \mathbf{k}\cdot\mathbf{v}_{i_0}= \frac{m}{r}v_{i_{\theta}} = m\varpi_0,
\end{equation}
\begin{equation}
\omega^{\star} = \mathbf{k}\cdot\mathbf{v}^{\star}= -\frac{m}{r}\frac{k_B T_e}{eB_0}\frac{{n_e}^{\prime}}{n_e}=-\frac{m}{r}k_{ne}\frac{k_B T_e}{eB_0},
\end{equation}
\begin{equation}
\omega_E = \mathbf{k}\cdot\mathbf{v}_{E \times B}= \frac{m}{r}\frac{{\phi_0}'}{B_0},
\end{equation}
\end{subequations}
where subscript $0$ indicates equilibrium quantities, and looking for a perturbation of the form
\begin{align}
\tilde{\phi} = & \phi_1(r)\exp\left[m\theta+k_z z -\omega t\right]\nonumber\\
\tilde{\textbf{v}}_i = & \textbf{v}_{i_1}(r)\exp\left[m\theta+k_z z -\omega t\right]\nonumber\\
\tilde{n}_i\ = & n_0(r) \eta_i(r)\left[m\theta+k_z z -\omega t\right],
\end{align}
the linearized ion momentum equation writes 
\begin{subequations}
\begin{equation}
-i(\omega-\omega_B) \tilde{v}_{i_r} -(2\varpi_0+\Omega_i)\tilde{v}_{i_{\theta}} = -\Omega_i \frac{\tilde{\phi}'}{B_0},
\end{equation}
\begin{equation}
(\Omega+2\varpi_0+r\varpi_0')\tilde{v}_{i_r} -i(\omega-\omega_B) \tilde{v}_{i_{\theta}}  = -i\Omega_i \frac{m}{r} \frac{\tilde{\phi}}{B_0},
\end{equation}
\begin{equation}
-i(\omega-\omega_B) \tilde{v}_{i_z} = -i\Omega_i k_z \frac{\tilde{\phi}}{B_0},
\end{equation}
\label{Eq:ion_linearized_momentum}
\end{subequations}
where $'$ denotes $\partial/\partial r$. The linearized ion continuity equation yields
 \begin{multline}
- i \omega \eta_i + {\tilde{v}_{i_r}}' + \frac{\tilde{v}_{i_r}}{r}+i \frac{m}{r}\tilde{v}_{i_{\theta}},+ i k_z \tilde{v}_{i_z} \\+ \tilde{v}_{i_r} \frac{{n_0}'}{n_0} + i\frac{m}{r}v_{i_{\theta}} \eta_i= 0.
 \end{multline}
Plugging in the ion velocity obtained from Eqs.~(\ref{Eq:ion_linearized_momentum}), and making use of the local approximation $\tilde{\phi}''=\tilde{\phi}'=0$ and $\tilde{n}''=\tilde{n}'=0$,  it rewrites
\begin{widetext}
\begin{multline}
 \eta_i = \chi \frac{{c_s}^2}{r^2(\omega-\omega_B)}\left[\frac{(k_z r)^2}{(\omega-\omega_B)}+\frac{m^2(\omega-\omega_B) + m \left[r{\varpi_0}' - (\Omega_i+2\varpi_0)\lambda\right]}{(\omega-\omega_B)^2-(2\varpi_0+\Omega_i)(2\varpi_0+r\varpi_0'+\Omega_i)} \right.\\+ \left.\frac{m(\Omega_i+2\varpi_0)\left[r{\varpi_0}'\left(5\left[\Omega+2\varpi_0\right]+2 r {\varpi_0}'\right)+r^2{\varpi_0}''(\Omega_i+2\varpi_0)\right]}{\left[(\omega-\omega_B)^2-\Omega_i^2-4\varpi_0(\Omega_i+\varpi_0)-r{\varpi_0}'(\Omega_i+2\varpi_0)\right]^2}\right],
 \label{Eq:ion_density_perturbation}
\end{multline}
\end{widetext}
with $\lambda = r |{n_0}'/n_0|$, ${c_s}^2 = k_B T_e/m_i$ and $\chi = e\tilde{\phi}/(k_B T_e)$. For solid body rotation equilibrium (${\varpi_0}''={\varpi_0}'=0$), the last term in the bracket on the right hand side in Eq.~(\ref{Eq:ion_density_perturbation}) cancels out, and the ion density - potential relation writes
\begin{multline}
\eta_i = \chi \frac{{c_s}^2}{r^2(\omega-\omega_B)}\left[\frac{(k_z r)^2}{(\omega-\omega_B)}\right.\\\left.+\frac{m^2(\omega-\omega_B) -m\sqrt{1-p}\Omega_i\lambda}{(\omega-\omega_B)^2-(1-p){\Omega_i}^2)} \right]
\label{Eq:ion_density_perturbation2}
\end{multline}
with $p = -4{\phi_0}^{\prime}/(r \Omega_i B_0)$. In the small $p$ ($p\sim 0$) and $(\omega-\omega_B)\gg\Omega_i$ limit, one recovers the usual expression
\begin{equation}
\eta_i = \chi \frac{{k}^2 {c_s}^2}{(\omega-\mathbf{k}\cdot\mathbf{v}_{i_0})^2}
\end{equation}
for non-magnetized ions.

Turning now to electrons, and looking similarly for a perturbation of the form
\begin{align}
\tilde{\textbf{v}}_e = & \textbf{v}_{e_1}(r)\exp\left[m\theta+k_z z -\omega t\right]\nonumber\\
\tilde{n}_e\ = & n_0(r) \eta_e(r)\left[m\theta+k_z z -\omega t\right],
\end{align}
the perpendicular components of the linearized electron momentum equation neglecting inertia and collisions writes
\begin{subequations}
\begin{equation}
\tilde{v}_{e_{\theta}} = \frac{{c_s}^2}{\Omega_i}\left(\chi'-{\eta_e}'\right),
\end{equation}
\begin{equation}
\tilde{v}_{e_r} = i \frac{m}{r} \frac{{c_s}^2}{\Omega_i}\left(\eta_e-\chi\right).
\end{equation}
In the parallel direction, one gets 
\begin{equation}
\tilde{v}_{e_z} = i k_z \frac{{v_{the}}^2}{\nu_{e}}\left(\chi-\eta_e\right),
\end{equation}
\label{Eq:electron_linearized_momentum}
\end{subequations}
with $\nu_e$ the electron collision frequency in the parallel direction. Plugging Eq.~(\ref{Eq:electron_linearized_momentum}) into the linearized continuity equation for electrons
\begin{multline}
-i \omega \eta_e+  {\tilde{v}_{e_r}}' + \frac{\tilde{v}_{e_r}}{r} + i \frac{m}{r} \tilde{v}_{e_{\theta}} + i k_z \tilde{v}_{e_z} \\+ \tilde{v}_{e_r} \frac{{n_0}'}{n_0} + i \frac{m}{r} v_{e_\theta} \eta_e = 0
\end{multline}
yields
\begin{equation}
\eta_e = \chi\frac{\displaystyle \omega^{\star}+i {k_z}^2\frac{\displaystyle {v_{the}}^2}{\displaystyle \nu_e}}{\displaystyle \omega-\omega_E+i {k_z}^2\frac{{v_{the}}^2}{\nu_e}},
\end{equation}
with $\omega^{\star} = -(m{c_s}^2{n_0}')/(r\Omega_i n_0) = m\lambda {c_s}^2/(r^2\Omega_i)$ and $\omega_E = (m{\phi_0}')/(rB_0)$. In the $\nu_e\sim 0$ and $k_z=0$ limit, we recover
\begin{equation}
\frac{\tilde{n}_e}{n_0} = \chi\frac{\displaystyle \omega^{\star}}{\displaystyle \omega-\omega_E}.
\label{Eq:electron_density_perturbation}
\end{equation}



Invoking quasi-neutrality $\eta_e=\eta_i$ for $k_z \sim 0$ and combining Eqs.~(\ref{Eq:ion_density_perturbation2}) and (\ref{Eq:electron_density_perturbation}), the complex frequency $\omega$ is found to be the solution of a cubic equation with real coefficients,  
\begin{multline}
\lambda^{-1}\left[m(\bar{\omega}+m\zeta/2)-\lambda(1-\zeta)\right]\\
\times \left[\bar{\omega}+m(1-(1-\zeta)^2)/4\right] \\
= (\bar{\omega}+m\zeta/2)^3-(1-\zeta)^2(\bar{\omega}+m\zeta/2),
\label{Eq:dispersion}
\end{multline}
with $\lambda = r |{n_0}'/n_0|$ and $\zeta = 1-\sqrt{1-p}$, and where frequencies are normalized to the ion cyclotron frequency, \emph{e.~g.} $\bar{\omega} = \omega/\Omega_i$. The stability criteria is thus equivalent to the positiveness of the discriminant of this cubic equation. Note that Eq.~(\ref{Eq:dispersion}) is equivalent to Eq.~(38) in Ref.~\cite{Chen1966} if replacing $N$ by
\begin{equation} 
N = m\frac{\uppsi(\uppsi^2-C^2)+C(\uppsi-m{\Upomega_0}^2)}{\uppsi(\uppsi-m{\Upomega_0}^2)},
\label{Eq:Chen}
\end{equation}
which corresponds to the negligible ion temperature ($T_i/T_e \sim 0$) and negligible plasma resistance limit. In Eq.~(\ref{Eq:Chen}), the normalized Doppler shifted frequency $\uppsi$, the normalized equilibrium ion angular frequency $\Upomega_0$ and the constant $C$ are consistent with Chen's notation~\cite{Chen1966}, 

In the limit $\lambda \Omega_i \ll m|(\omega-\omega_B)|$, \emph{i.~e.} weak density gradient, Eq.~(\ref{Eq:dispersion}) reduces to a quadratic equation, of which the roots are
\begin{multline}
\bar{\omega}_{1,2} = \frac{m(\lambda^{-1}-\zeta)}{2}\\ \pm\sqrt{\frac{m}{\lambda}\left[\frac{m}{4}(\lambda^{-1}-\zeta^2)\right]+(1-\zeta)^2}
\label{Eq:sol_second_order}
\end{multline}
We note that the bracketed term under the square root is equal to ${c_s}^2m^2/(4r^2\omega^{\star}\Omega_i)+(\bar{\omega}_b-\bar{\omega}_E)$. The corresponding stability region is plotted in Fig.~\ref{Fig:Simple}. In the limit $\lambda\ll 1$, and since $0\leq\zeta\leq1$, this further simplifies to give $\omega = 0$ or $\omega = m/\lambda$.

\begin{figure}
\includegraphics{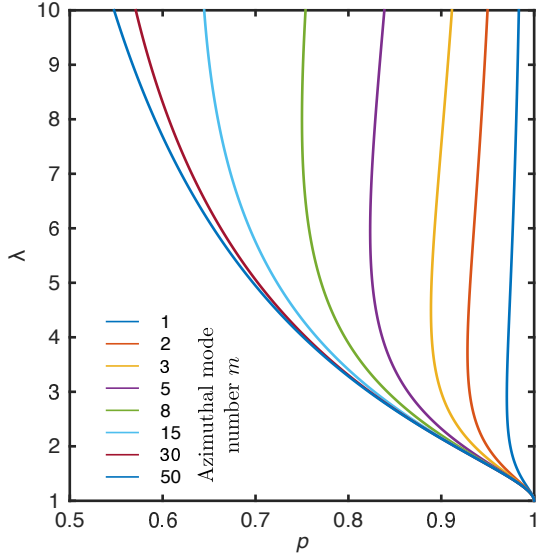}
\caption{Instability threshold resulting from fast rotation as a function of the Brillouin parameter $p = -4{\phi_0}^{\prime}/(r \Omega_i B_0)$ and the dimensionless density gradient scale-length $\lambda = r |{n_0}'/n_0|$ for various mode numbers $m$. The region to the left of a given line is stable, while the region to the right of the same line is unstable.}
\label{Fig:Simple}
\end{figure}


\begin{thebibliography}{47}%
\makeatletter
\providecommand \@ifxundefined [1]{%
 \@ifx{#1\undefined}
}%
\providecommand \@ifnum [1]{%
 \ifnum #1\expandafter \@firstoftwo
 \else \expandafter \@secondoftwo
 \fi
}%
\providecommand \@ifx [1]{%
 \ifx #1\expandafter \@firstoftwo
 \else \expandafter \@secondoftwo
 \fi
}%
\providecommand \natexlab [1]{#1}%
\providecommand \enquote  [1]{``#1''}%
\providecommand \bibnamefont  [1]{#1}%
\providecommand \bibfnamefont [1]{#1}%
\providecommand \citenamefont [1]{#1}%
\providecommand \href@noop [0]{\@secondoftwo}%
\providecommand \href [0]{\begingroup \@sanitize@url \@href}%
\providecommand \@href[1]{\@@startlink{#1}\@@href}%
\providecommand \@@href[1]{\endgroup#1\@@endlink}%
\providecommand \@sanitize@url [0]{\catcode `\\12\catcode `\$12\catcode
  `\&12\catcode `\#12\catcode `\^12\catcode `\_12\catcode `\%12\relax}%
\providecommand \@@startlink[1]{}%
\providecommand \@@endlink[0]{}%
\providecommand \url  [0]{\begingroup\@sanitize@url \@url }%
\providecommand \@url [1]{\endgroup\@href {#1}{\urlprefix }}%
\providecommand \urlprefix  [0]{URL }%
\providecommand \Eprint [0]{\href }%
\providecommand \doibase [0]{http://dx.doi.org/}%
\providecommand \selectlanguage [0]{\@gobble}%
\providecommand \bibinfo  [0]{\@secondoftwo}%
\providecommand \bibfield  [0]{\@secondoftwo}%
\providecommand \translation [1]{[#1]}%
\providecommand \BibitemOpen [0]{}%
\providecommand \bibitemStop [0]{}%
\providecommand \bibitemNoStop [0]{.\EOS\space}%
\providecommand \EOS [0]{\spacefactor3000\relax}%
\providecommand \BibitemShut  [1]{\csname bibitem#1\endcsname}%
\let\auto@bib@innerbib\@empty
\bibitem [{\citenamefont {Tsuda}\ \emph {et~al.}(1969)\citenamefont {Tsuda},
  \citenamefont {Sato},\ and\ \citenamefont {Matsushita}}]{Tsuda1969}%
  \BibitemOpen
  \bibfield  {author} {\bibinfo {author} {\bibfnamefont {T.}~\bibnamefont
  {Tsuda}}, \bibinfo {author} {\bibfnamefont {T.}~\bibnamefont {Sato}}, \ and\
  \bibinfo {author} {\bibfnamefont {S.}~\bibnamefont {Matsushita}},\ }\href
  {\doibase 10.1029/JA074i011p02923} {\bibfield  {journal} {\bibinfo  {journal}
  {J. Geophys. Res.}\ } (\bibinfo {year} {1969}),\
  10.1029/JA074i011p02923}\BibitemShut {NoStop}%
\bibitem [{\citenamefont {Taylor}\ \emph {et~al.}(1989)\citenamefont {Taylor},
  \citenamefont {Brown}, \citenamefont {Fried}, \citenamefont {Grote},
  \citenamefont {Liberati}, \citenamefont {Morales}, \citenamefont {Pribyl},
  \citenamefont {Darrow},\ and\ \citenamefont {Ono}}]{Taylor1989}%
  \BibitemOpen
  \bibfield  {author} {\bibinfo {author} {\bibfnamefont {R.~J.}\ \bibnamefont
  {Taylor}}, \bibinfo {author} {\bibfnamefont {M.~L.}\ \bibnamefont {Brown}},
  \bibinfo {author} {\bibfnamefont {B.~D.}\ \bibnamefont {Fried}}, \bibinfo
  {author} {\bibfnamefont {H.}~\bibnamefont {Grote}}, \bibinfo {author}
  {\bibfnamefont {J.~R.}\ \bibnamefont {Liberati}}, \bibinfo {author}
  {\bibfnamefont {G.~J.}\ \bibnamefont {Morales}}, \bibinfo {author}
  {\bibfnamefont {P.}~\bibnamefont {Pribyl}}, \bibinfo {author} {\bibfnamefont
  {D.}~\bibnamefont {Darrow}}, \ and\ \bibinfo {author} {\bibfnamefont
  {M.}~\bibnamefont {Ono}},\ }\href {\doibase
  https://doi.org/10.1103/PhysRevLett.63.2365} {\bibfield  {journal} {\bibinfo
  {journal} {Phys. Rev. Lett.}\ }\textbf {\bibinfo {volume} {63}},\ \bibinfo
  {pages} {2365} (\bibinfo {year} {1989})}\BibitemShut {NoStop}%
\bibitem [{\citenamefont {Abolmasov}(2012)}]{Abolmasov2012}%
  \BibitemOpen
  \bibfield  {author} {\bibinfo {author} {\bibfnamefont {S.~N.}\ \bibnamefont
  {Abolmasov}},\ }\href {\doibase 10.1088/0963-0252/21/3/035006} {\bibfield
  {journal} {\bibinfo  {journal} {Plasma Sources Sci. Technol.}\ }\textbf
  {\bibinfo {volume} {21}},\ \bibinfo {pages} {035006} (\bibinfo {year}
  {2012})}\BibitemShut {NoStop}%
\bibitem [{\citenamefont {Morozov}\ and\ \citenamefont
  {Savelyev}(2000)}]{Morozov2000}%
  \BibitemOpen
  \bibfield  {author} {\bibinfo {author} {\bibfnamefont {A.~I.}\ \bibnamefont
  {Morozov}}\ and\ \bibinfo {author} {\bibfnamefont {V.~V.}\ \bibnamefont
  {Savelyev}},\ }in\ \href {\doibase 10.1007/978-1-4615-4309-1_2} {\emph
  {\bibinfo {booktitle} {Reviews of Plasma Physics}}},\ Vol.~\bibinfo {volume}
  {21},\ \bibinfo {editor} {edited by\ \bibinfo {editor} {\bibfnamefont
  {B.~B.}\ \bibnamefont {Kadomtsev}}\ and\ \bibinfo {editor} {\bibfnamefont
  {V.~D.}\ \bibnamefont {Shafranov}}}\ (\bibinfo {year} {2000})\ p.\ \bibinfo
  {pages} {203}\BibitemShut {NoStop}%
\bibitem [{\citenamefont {Boeuf}(2017)}]{Boeuf2017}%
  \BibitemOpen
  \bibfield  {author} {\bibinfo {author} {\bibfnamefont {J.-P.}\ \bibnamefont
  {Boeuf}},\ }\href {\doibase 10.1063/1.4972269} {\bibfield  {journal}
  {\bibinfo  {journal} {J. Appl. Phys.}\ }\textbf {\bibinfo {volume} {121}},\
  \bibinfo {pages} {011101} (\bibinfo {year} {2017})}\BibitemShut {NoStop}%
\bibitem [{\citenamefont {Bonnevier}(1966)}]{Bonnevier1966}%
  \BibitemOpen
  \bibfield  {author} {\bibinfo {author} {\bibfnamefont {B.}~\bibnamefont
  {Bonnevier}},\ }\href@noop {} {\bibfield  {journal} {\bibinfo  {journal}
  {Ark. Fys.}\ }\textbf {\bibinfo {volume} {33}},\ \bibinfo {pages} {255}
  (\bibinfo {year} {1966})}\BibitemShut {NoStop}%
\bibitem [{\citenamefont {Lehnert}(1971)}]{Lehnert1971}%
  \BibitemOpen
  \bibfield  {author} {\bibinfo {author} {\bibfnamefont {B.}~\bibnamefont
  {Lehnert}},\ }\href {\doibase 10.1088/0029-5515/11/5/010} {\bibfield
  {journal} {\bibinfo  {journal} {Nucl. Fusion}\ }\textbf {\bibinfo {volume}
  {11}},\ \bibinfo {pages} {485} (\bibinfo {year} {1971})}\BibitemShut
  {NoStop}%
\bibitem [{\citenamefont {Prasad}\ and\ \citenamefont
  {Krishnan}(1987)}]{Prasad1987}%
  \BibitemOpen
  \bibfield  {author} {\bibinfo {author} {\bibfnamefont {R.~R.}\ \bibnamefont
  {Prasad}}\ and\ \bibinfo {author} {\bibfnamefont {M.}~\bibnamefont
  {Krishnan}},\ }\href {\doibase 10.1063/1.338407} {\bibfield  {journal}
  {\bibinfo  {journal} {Journal of Applied Physics}\ }\textbf {\bibinfo
  {volume} {61}},\ \bibinfo {pages} {4464} (\bibinfo {year}
  {1987})}\BibitemShut {NoStop}%
\bibitem [{\citenamefont {Ohkawa}\ and\ \citenamefont
  {Miller}(2002)}]{Ohkawa2002}%
  \BibitemOpen
  \bibfield  {author} {\bibinfo {author} {\bibfnamefont {T.}~\bibnamefont
  {Ohkawa}}\ and\ \bibinfo {author} {\bibfnamefont {R.~L.}\ \bibnamefont
  {Miller}},\ }\href {\doibase 10.1063/1.1523930} {\bibfield  {journal}
  {\bibinfo  {journal} {Phys. Plasmas}\ }\textbf {\bibinfo {volume} {9}},\
  \bibinfo {pages} {5116} (\bibinfo {year} {2002})}\BibitemShut {NoStop}%
\bibitem [{\citenamefont {Fetterman}\ and\ \citenamefont
  {Fisch}(2011)}]{Fetterman2011}%
  \BibitemOpen
  \bibfield  {author} {\bibinfo {author} {\bibfnamefont {A.~J.}\ \bibnamefont
  {Fetterman}}\ and\ \bibinfo {author} {\bibfnamefont {N.~J.}\ \bibnamefont
  {Fisch}},\ }\href {\doibase 10.1063/1.3631793} {\bibfield  {journal}
  {\bibinfo  {journal} {Phys. Plasmas}\ }\textbf {\bibinfo {volume} {18}},\
  \bibinfo {pages} {094503} (\bibinfo {year} {2011})}\BibitemShut {NoStop}%
\bibitem [{\citenamefont {Gueroult}\ \emph {et~al.}(2014)\citenamefont
  {Gueroult}, \citenamefont {Rax},\ and\ \citenamefont {Fisch}}]{Gueroult2014}%
  \BibitemOpen
  \bibfield  {author} {\bibinfo {author} {\bibfnamefont {R.}~\bibnamefont
  {Gueroult}}, \bibinfo {author} {\bibfnamefont {J.-M.}\ \bibnamefont {Rax}}, \
  and\ \bibinfo {author} {\bibfnamefont {N.~J.}\ \bibnamefont {Fisch}},\ }\href
  {\doibase 10.1063/1.4864325} {\bibfield  {journal} {\bibinfo  {journal}
  {Phys. Plasmas}\ }\textbf {\bibinfo {volume} {21}},\ \bibinfo {pages}
  {020701} (\bibinfo {year} {2014})}\BibitemShut {NoStop}%
\bibitem [{\citenamefont {Rax}\ and\ \citenamefont {Gueroult}(2016)}]{Rax2016}%
  \BibitemOpen
  \bibfield  {author} {\bibinfo {author} {\bibfnamefont {J.-M.}\ \bibnamefont
  {Rax}}\ and\ \bibinfo {author} {\bibfnamefont {R.}~\bibnamefont {Gueroult}},\
  }\href {\doibase 10.1017/S0022377816000878} {\bibfield  {journal} {\bibinfo
  {journal} {J. Plasma Phys.}\ }\textbf {\bibinfo {volume} {82}},\ \bibinfo
  {pages} {595820504} (\bibinfo {year} {2016})}\BibitemShut {NoStop}%
\bibitem [{\citenamefont {Geyko}\ and\ \citenamefont
  {Fisch}(2017)}]{Geyko2017}%
  \BibitemOpen
  \bibfield  {author} {\bibinfo {author} {\bibfnamefont {V.~I.}\ \bibnamefont
  {Geyko}}\ and\ \bibinfo {author} {\bibfnamefont {N.~J.}\ \bibnamefont
  {Fisch}},\ }\href {\doibase 10.1063/1.4975651} {\bibfield  {journal}
  {\bibinfo  {journal} {Phys. Plasmas}\ }\textbf {\bibinfo {volume} {24}},\
  \bibinfo {pages} {022113} (\bibinfo {year} {2017})}\BibitemShut {NoStop}%
\bibitem [{\citenamefont {Rax}\ \emph {et~al.}(2017)\citenamefont {Rax},
  \citenamefont {Gueroult},\ and\ \citenamefont {Fisch}}]{Rax2017}%
  \BibitemOpen
  \bibfield  {author} {\bibinfo {author} {\bibfnamefont {J.~M.}\ \bibnamefont
  {Rax}}, \bibinfo {author} {\bibfnamefont {R.}~\bibnamefont {Gueroult}}, \
  and\ \bibinfo {author} {\bibfnamefont {N.~J.}\ \bibnamefont {Fisch}},\
  }\bibfield  {booktitle} {\emph {\bibinfo {booktitle} {Physics of Plasmas}},\
  }\href {\doibase 10.1063/1.4977919} {\bibfield  {journal} {\bibinfo
  {journal} {Phys. Plasmas}\ }\textbf {\bibinfo {volume} {24}},\ \bibinfo
  {pages} {032504} (\bibinfo {year} {2017})}\BibitemShut {NoStop}%
\bibitem [{\citenamefont {Redhead}(1988)}]{Redhead1988}%
  \BibitemOpen
  \bibfield  {author} {\bibinfo {author} {\bibfnamefont {P.~A.}\ \bibnamefont
  {Redhead}},\ }\href {\doibase 10.1016/0042-207X(88)90489-7} {\bibfield
  {journal} {\bibinfo  {journal} {Vacuum}\ }\textbf {\bibinfo {volume} {38}},\
  \bibinfo {pages} {901} (\bibinfo {year} {1988})}\BibitemShut {NoStop}%
\bibitem [{\citenamefont {Horton}(1999)}]{Horton1999}%
  \BibitemOpen
  \bibfield  {author} {\bibinfo {author} {\bibfnamefont {W.}~\bibnamefont
  {Horton}},\ }\href {\doibase 10.1103/RevModPhys.71.735} {\bibfield  {journal}
  {\bibinfo  {journal} {Rev. Mod. Phys.}\ }\textbf {\bibinfo {volume} {71}},\
  \bibinfo {pages} {735} (\bibinfo {year} {1999})}\BibitemShut {NoStop}%
\bibitem [{\citenamefont {Litvak}\ and\ \citenamefont
  {Fisch}(2001)}]{Litvak2001}%
  \BibitemOpen
  \bibfield  {author} {\bibinfo {author} {\bibfnamefont {A.~A.}\ \bibnamefont
  {Litvak}}\ and\ \bibinfo {author} {\bibfnamefont {N.~J.}\ \bibnamefont
  {Fisch}},\ }\href {\doibase 10.1063/1.1336531} {\bibfield  {journal}
  {\bibinfo  {journal} {Phys. Plasmas}\ }\textbf {\bibinfo {volume} {8}},\
  \bibinfo {pages} {648} (\bibinfo {year} {2001})}\BibitemShut {NoStop}%
\bibitem [{\citenamefont {Litvak}\ and\ \citenamefont
  {Fisch}(2004)}]{Litvak2004}%
  \BibitemOpen
  \bibfield  {author} {\bibinfo {author} {\bibfnamefont {A.~A.}\ \bibnamefont
  {Litvak}}\ and\ \bibinfo {author} {\bibfnamefont {N.~J.}\ \bibnamefont
  {Fisch}},\ }\href {\doibase 10.1063/1.1647565} {\bibfield  {journal}
  {\bibinfo  {journal} {Phys. Plasmas}\ }\textbf {\bibinfo {volume} {11}},\
  \bibinfo {pages} {1379} (\bibinfo {year} {2004})}\BibitemShut {NoStop}%
\bibitem [{\citenamefont {Smolyakov}\ \emph {et~al.}(2017)\citenamefont
  {Smolyakov}, \citenamefont {Chapurin}, \citenamefont {Frias}, \citenamefont
  {Koshkarov}, \citenamefont {Romadanov}, \citenamefont {Tang}, \citenamefont
  {Umansky}, \citenamefont {Raitses}, \citenamefont {Kaganovich},\ and\
  \citenamefont {Lakhin}}]{Smolyakov2017}%
  \BibitemOpen
  \bibfield  {author} {\bibinfo {author} {\bibfnamefont {A.~I.}\ \bibnamefont
  {Smolyakov}}, \bibinfo {author} {\bibfnamefont {O.}~\bibnamefont {Chapurin}},
  \bibinfo {author} {\bibfnamefont {W.}~\bibnamefont {Frias}}, \bibinfo
  {author} {\bibfnamefont {O.}~\bibnamefont {Koshkarov}}, \bibinfo {author}
  {\bibfnamefont {I.}~\bibnamefont {Romadanov}}, \bibinfo {author}
  {\bibfnamefont {T.}~\bibnamefont {Tang}}, \bibinfo {author} {\bibfnamefont
  {M.}~\bibnamefont {Umansky}}, \bibinfo {author} {\bibfnamefont
  {Y.}~\bibnamefont {Raitses}}, \bibinfo {author} {\bibfnamefont {I.~D.}\
  \bibnamefont {Kaganovich}}, \ and\ \bibinfo {author} {\bibfnamefont {V.~P.}\
  \bibnamefont {Lakhin}},\ }\href {\doibase 10.1088/0741-3335/59/1/014041}
  {\bibfield  {journal} {\bibinfo  {journal} {Plasma Phys. Controlled Fusion}\
  }\textbf {\bibinfo {volume} {59}},\ \bibinfo {pages} {014041} (\bibinfo
  {year} {2017})}\BibitemShut {NoStop}%
\bibitem [{\citenamefont {Simon}(1963)}]{Simon1963}%
  \BibitemOpen
  \bibfield  {author} {\bibinfo {author} {\bibfnamefont {A.}~\bibnamefont
  {Simon}},\ }\href {\doibase 10.1063/1.1706743} {\bibfield  {journal}
  {\bibinfo  {journal} {Phys. Fluids}\ }\textbf {\bibinfo {volume} {6}},\
  \bibinfo {pages} {382} (\bibinfo {year} {1963})}\BibitemShut {NoStop}%
\bibitem [{\citenamefont {Hoh}(1963{\natexlab{a}})}]{Hoh1963}%
  \BibitemOpen
  \bibfield  {author} {\bibinfo {author} {\bibfnamefont {F.~C.}\ \bibnamefont
  {Hoh}},\ }\href {\doibase 10.1063/1.1706878} {\bibfield  {journal} {\bibinfo
  {journal} {Phys. Fluids}\ }\textbf {\bibinfo {volume} {6}},\ \bibinfo {pages}
  {1184} (\bibinfo {year} {1963}{\natexlab{a}})}\BibitemShut {NoStop}%
\bibitem [{\citenamefont {Sakawa}\ \emph {et~al.}(1993)\citenamefont {Sakawa},
  \citenamefont {Joshi}, \citenamefont {Kaw}, \citenamefont {Chen},\ and\
  \citenamefont {Jain}}]{Sakawa1993}%
  \BibitemOpen
  \bibfield  {author} {\bibinfo {author} {\bibfnamefont {Y.}~\bibnamefont
  {Sakawa}}, \bibinfo {author} {\bibfnamefont {C.}~\bibnamefont {Joshi}},
  \bibinfo {author} {\bibfnamefont {P.~K.}\ \bibnamefont {Kaw}}, \bibinfo
  {author} {\bibfnamefont {F.~F.}\ \bibnamefont {Chen}}, \ and\ \bibinfo
  {author} {\bibfnamefont {V.~K.}\ \bibnamefont {Jain}},\ }\href {\doibase
  10.1063/1.860803} {\bibfield  {journal} {\bibinfo  {journal} {Phys. Fluids B:
  Plasma Phys.}\ }\textbf {\bibinfo {volume} {5}},\ \bibinfo {pages} {1681}
  (\bibinfo {year} {1993})}\BibitemShut {NoStop}%
\bibitem [{\citenamefont {Sakawa}\ \emph {et~al.}(1992)\citenamefont {Sakawa},
  \citenamefont {Joshi}, \citenamefont {Kaw}, \citenamefont {Jain},
  \citenamefont {Johnston}, \citenamefont {Chen},\ and\ \citenamefont
  {Dawson}}]{Sakawa1992}%
  \BibitemOpen
  \bibfield  {author} {\bibinfo {author} {\bibfnamefont {Y.}~\bibnamefont
  {Sakawa}}, \bibinfo {author} {\bibfnamefont {C.}~\bibnamefont {Joshi}},
  \bibinfo {author} {\bibfnamefont {P.~K.}\ \bibnamefont {Kaw}}, \bibinfo
  {author} {\bibfnamefont {V.~K.}\ \bibnamefont {Jain}}, \bibinfo {author}
  {\bibfnamefont {T.~W.}\ \bibnamefont {Johnston}}, \bibinfo {author}
  {\bibfnamefont {F.~F.}\ \bibnamefont {Chen}}, \ and\ \bibinfo {author}
  {\bibfnamefont {J.~M.}\ \bibnamefont {Dawson}},\ }\href {\doibase
  10.1103/PhysRevLett.69.85} {\bibfield  {journal} {\bibinfo  {journal} {Phys.
  Rev. Lett.}\ }\textbf {\bibinfo {volume} {69}},\ \bibinfo {pages} {85}
  (\bibinfo {year} {1992})}\BibitemShut {NoStop}%
\bibitem [{\citenamefont {Sakawa}\ and\ \citenamefont
  {Joshi}(2000)}]{Sakawa2000}%
  \BibitemOpen
  \bibfield  {author} {\bibinfo {author} {\bibfnamefont {Y.}~\bibnamefont
  {Sakawa}}\ and\ \bibinfo {author} {\bibfnamefont {C.}~\bibnamefont {Joshi}},\
  }\href {\doibase 10.1063/1.873998} {\bibfield  {journal} {\bibinfo  {journal}
  {Phys. Plasmas}\ }\textbf {\bibinfo {volume} {7}},\ \bibinfo {pages} {1774}
  (\bibinfo {year} {2000})}\BibitemShut {NoStop}%
\bibitem [{\citenamefont {Frias}\ \emph {et~al.}(2012)\citenamefont {Frias},
  \citenamefont {Smolyakov}, \citenamefont {Kaganovich},\ and\ \citenamefont
  {Raitses}}]{Frias2012}%
  \BibitemOpen
  \bibfield  {author} {\bibinfo {author} {\bibfnamefont {W.}~\bibnamefont
  {Frias}}, \bibinfo {author} {\bibfnamefont {A.~I.}\ \bibnamefont
  {Smolyakov}}, \bibinfo {author} {\bibfnamefont {I.~D.}\ \bibnamefont
  {Kaganovich}}, \ and\ \bibinfo {author} {\bibfnamefont {Y.}~\bibnamefont
  {Raitses}},\ }\href {\doibase 10.1063/1.4736997} {\bibfield  {journal}
  {\bibinfo  {journal} {Phys. Plasmas}\ }\textbf {\bibinfo {volume} {19}},\
  \bibinfo {pages} {072112} (\bibinfo {year} {2012})}\BibitemShut {NoStop}%
\bibitem [{\citenamefont {Frias}\ \emph {et~al.}(2013)\citenamefont {Frias},
  \citenamefont {Smolyakov}, \citenamefont {Kaganovich},\ and\ \citenamefont
  {Raitses}}]{Frias2013}%
  \BibitemOpen
  \bibfield  {author} {\bibinfo {author} {\bibfnamefont {W.}~\bibnamefont
  {Frias}}, \bibinfo {author} {\bibfnamefont {A.~I.}\ \bibnamefont
  {Smolyakov}}, \bibinfo {author} {\bibfnamefont {I.~D.}\ \bibnamefont
  {Kaganovich}}, \ and\ \bibinfo {author} {\bibfnamefont {Y.}~\bibnamefont
  {Raitses}},\ }\href {\doibase 10.1063/1.4804281} {\bibfield  {journal}
  {\bibinfo  {journal} {Phys. Plasmas}\ }\textbf {\bibinfo {volume} {20}},\
  \bibinfo {pages} {052108} (\bibinfo {year} {2013})}\BibitemShut {NoStop}%
\bibitem [{\citenamefont {Fridman}(1964)}]{Fridman1964}%
  \BibitemOpen
  \bibfield  {author} {\bibinfo {author} {\bibfnamefont {A.~M.}\ \bibnamefont
  {Fridman}},\ }\href@noop {} {\bibfield  {journal} {\bibinfo  {journal} {Sov.
  Phys. Dokl.}\ }\textbf {\bibinfo {volume} {9}},\ \bibinfo {pages} {75}
  (\bibinfo {year} {1964})}\BibitemShut {NoStop}%
\bibitem [{\citenamefont {Rosenbluth}\ \emph {et~al.}(1962)\citenamefont
  {Rosenbluth}, \citenamefont {Krall},\ and\ \citenamefont
  {Rostoker}}]{Rosenbluth1962}%
  \BibitemOpen
  \bibfield  {author} {\bibinfo {author} {\bibfnamefont {M.~N.}\ \bibnamefont
  {Rosenbluth}}, \bibinfo {author} {\bibfnamefont {N.~A.}\ \bibnamefont
  {Krall}}, \ and\ \bibinfo {author} {\bibfnamefont {N.}~\bibnamefont
  {Rostoker}},\ }\href@noop {} {\bibfield  {journal} {\bibinfo  {journal}
  {Nucl. Fusion Suppl. Pt. 1}\ ,\ \bibinfo {pages} {143}} (\bibinfo {year}
  {1962})}\BibitemShut {NoStop}%
\bibitem [{\citenamefont {Rosenbluth}\ and\ \citenamefont
  {Longmire}(1957)}]{Rosenbluth1957}%
  \BibitemOpen
  \bibfield  {author} {\bibinfo {author} {\bibfnamefont {M.~N.}\ \bibnamefont
  {Rosenbluth}}\ and\ \bibinfo {author} {\bibfnamefont {C.~L.}\ \bibnamefont
  {Longmire}},\ }\href {\doibase 10.1016/0003-4916(57)90055-6} {\bibfield
  {journal} {\bibinfo  {journal} {Ann. Phys.}\ }\textbf {\bibinfo {volume}
  {1}},\ \bibinfo {pages} {120} (\bibinfo {year} {1957})}\BibitemShut {NoStop}%
\bibitem [{\citenamefont {Chen}(1966)}]{Chen1966}%
  \BibitemOpen
  \bibfield  {author} {\bibinfo {author} {\bibfnamefont {F.~F.}\ \bibnamefont
  {Chen}},\ }\href {\doibase 10.1063/1.1761798} {\bibfield  {journal} {\bibinfo
   {journal} {Phys. Fluids}\ }\textbf {\bibinfo {volume} {9}},\ \bibinfo
  {pages} {965} (\bibinfo {year} {1966})}\BibitemShut {NoStop}%
\bibitem [{\citenamefont {Chen}(1967)}]{Chen1967}%
  \BibitemOpen
  \bibfield  {author} {\bibinfo {author} {\bibfnamefont {F.~F.}\ \bibnamefont
  {Chen}},\ }\href {\doibase 10.1063/1.1762340} {\bibfield  {journal} {\bibinfo
   {journal} {Phys. Fluids}\ }\textbf {\bibinfo {volume} {10}},\ \bibinfo
  {pages} {1647} (\bibinfo {year} {1967})}\BibitemShut {NoStop}%
\bibitem [{\citenamefont {Rognlien}(1973)}]{Rognlien1973}%
  \BibitemOpen
  \bibfield  {author} {\bibinfo {author} {\bibfnamefont {T.~D.}\ \bibnamefont
  {Rognlien}},\ }\href {\doibase 10.1063/1.1662794} {\bibfield  {journal}
  {\bibinfo  {journal} {J. Appl. Phys.}\ }\textbf {\bibinfo {volume} {44}},\
  \bibinfo {pages} {3505} (\bibinfo {year} {1973})}\BibitemShut {NoStop}%
\bibitem [{\citenamefont {Rosenbluth}\ and\ \citenamefont
  {Simon}(1965)}]{Rosenbluth1965}%
  \BibitemOpen
  \bibfield  {author} {\bibinfo {author} {\bibfnamefont {M.~N.}\ \bibnamefont
  {Rosenbluth}}\ and\ \bibinfo {author} {\bibfnamefont {A.}~\bibnamefont
  {Simon}},\ }\href {\doibase 10.1063/1.1761402} {\bibfield  {journal}
  {\bibinfo  {journal} {Phys. Fluids}\ }\textbf {\bibinfo {volume} {8}},\
  \bibinfo {pages} {1300} (\bibinfo {year} {1965})}\BibitemShut {NoStop}%
\bibitem [{\citenamefont {Perkins}\ and\ \citenamefont
  {Jassby}(1971)}]{Perkins1971}%
  \BibitemOpen
  \bibfield  {author} {\bibinfo {author} {\bibfnamefont {F.~W.}\ \bibnamefont
  {Perkins}}\ and\ \bibinfo {author} {\bibfnamefont {D.~L.}\ \bibnamefont
  {Jassby}},\ }\href {\doibase 10.1063/1.1693259} {\bibfield  {journal}
  {\bibinfo  {journal} {Phys. Fluids}\ }\textbf {\bibinfo {volume} {14}},\
  \bibinfo {pages} {102} (\bibinfo {year} {1971})}\BibitemShut {NoStop}%
\bibitem [{\citenamefont {Jassby}(1972)}]{Jassby1972}%
  \BibitemOpen
  \bibfield  {author} {\bibinfo {author} {\bibfnamefont {D.~L.}\ \bibnamefont
  {Jassby}},\ }\href {\doibase 10.1063/1.1694135} {\bibfield  {journal}
  {\bibinfo  {journal} {Phys. Fluids}\ }\textbf {\bibinfo {volume} {15}},\
  \bibinfo {pages} {1590} (\bibinfo {year} {1972})}\BibitemShut {NoStop}%
\bibitem [{\citenamefont {Simon}\ and\ \citenamefont
  {Rosenbluth}(1966)}]{Simon1966}%
  \BibitemOpen
  \bibfield  {author} {\bibinfo {author} {\bibfnamefont {A.}~\bibnamefont
  {Simon}}\ and\ \bibinfo {author} {\bibfnamefont {M.~N.}\ \bibnamefont
  {Rosenbluth}},\ }\href {\doibase 10.1063/1.1761739} {\bibfield  {journal}
  {\bibinfo  {journal} {Phys. Fluids}\ }\textbf {\bibinfo {volume} {9}},\
  \bibinfo {pages} {726} (\bibinfo {year} {1966})}\BibitemShut {NoStop}%
\bibitem [{\citenamefont {Hole}\ \emph {et~al.}(2002)\citenamefont {Hole},
  \citenamefont {Dallaqua}, \citenamefont {Simpson},\ and\ \citenamefont
  {Del~Bosco}}]{Hole2002}%
  \BibitemOpen
  \bibfield  {author} {\bibinfo {author} {\bibfnamefont {M.~J.}\ \bibnamefont
  {Hole}}, \bibinfo {author} {\bibfnamefont {R.~S.}\ \bibnamefont {Dallaqua}},
  \bibinfo {author} {\bibfnamefont {S.~W.}\ \bibnamefont {Simpson}}, \ and\
  \bibinfo {author} {\bibfnamefont {E.}~\bibnamefont {Del~Bosco}},\ }\href
  {\doibase 10.1103/PhysRevE.65.046409} {\bibfield  {journal} {\bibinfo
  {journal} {Phys. Rev. E}\ }\textbf {\bibinfo {volume} {65}},\ \bibinfo
  {pages} {046409} (\bibinfo {year} {2002})}\BibitemShut {NoStop}%
\bibitem [{\citenamefont {Ilic}\ \emph {et~al.}(1973)\citenamefont {Ilic},
  \citenamefont {Rognlien}, \citenamefont {Self},\ and\ \citenamefont
  {Crawford}}]{Ilic1973}%
  \BibitemOpen
  \bibfield  {author} {\bibinfo {author} {\bibfnamefont {D.~B.}\ \bibnamefont
  {Ilic}}, \bibinfo {author} {\bibfnamefont {T.~D.}\ \bibnamefont {Rognlien}},
  \bibinfo {author} {\bibfnamefont {S.~A.}\ \bibnamefont {Self}}, \ and\
  \bibinfo {author} {\bibfnamefont {F.~W.}\ \bibnamefont {Crawford}},\ }\href
  {\doibase 10.1063/1.1694466} {\bibfield  {journal} {\bibinfo  {journal}
  {Phys. Fluids}\ }\textbf {\bibinfo {volume} {16}},\ \bibinfo {pages} {1042}
  (\bibinfo {year} {1973})}\BibitemShut {NoStop}%
\bibitem [{\citenamefont {Davidson}(2001)}]{Davidson2001}%
  \BibitemOpen
  \bibfield  {author} {\bibinfo {author} {\bibfnamefont {R.~C.}\ \bibnamefont
  {Davidson}},\ }\href@noop {} {\emph {\bibinfo {title} {Physics of Nonneutral
  Plasmas}}}\ (\bibinfo  {publisher} {Imperial College Press},\ \bibinfo {year}
  {2001})\BibitemShut {NoStop}%
\bibitem [{\citenamefont {Rax}\ \emph {et~al.}(2015)\citenamefont {Rax},
  \citenamefont {Fruchtman}, \citenamefont {Gueroult},\ and\ \citenamefont
  {Fisch}}]{Rax2015}%
  \BibitemOpen
  \bibfield  {author} {\bibinfo {author} {\bibfnamefont {J.~M.}\ \bibnamefont
  {Rax}}, \bibinfo {author} {\bibfnamefont {A.}~\bibnamefont {Fruchtman}},
  \bibinfo {author} {\bibfnamefont {R.}~\bibnamefont {Gueroult}}, \ and\
  \bibinfo {author} {\bibfnamefont {N.~J.}\ \bibnamefont {Fisch}},\ }\href
  {\doibase 10.1063/1.4929791} {\bibfield  {journal} {\bibinfo  {journal}
  {Phys. Plasmas}\ }\textbf {\bibinfo {volume} {22}},\ \bibinfo {pages}
  {092101} (\bibinfo {year} {2015})}\BibitemShut {NoStop}%
\bibitem [{\citenamefont {Hoh}(1963{\natexlab{b}})}]{Hoh1963a}%
  \BibitemOpen
  \bibfield  {author} {\bibinfo {author} {\bibfnamefont {F.~C.}\ \bibnamefont
  {Hoh}},\ }\href {\doibase 10.1063/1.1706909} {\bibfield  {journal} {\bibinfo
  {journal} {Phys. Fluids}\ }\textbf {\bibinfo {volume} {6}},\ \bibinfo {pages}
  {1359} (\bibinfo {year} {1963}{\natexlab{b}})}\BibitemShut {NoStop}%
\bibitem [{\citenamefont {Shinohara}\ and\ \citenamefont
  {Horii}(2007)}]{Shinohara2007}%
  \BibitemOpen
  \bibfield  {author} {\bibinfo {author} {\bibfnamefont {S.}~\bibnamefont
  {Shinohara}}\ and\ \bibinfo {author} {\bibfnamefont {S.}~\bibnamefont
  {Horii}},\ }\href {\doibase 10.1143/JJAP.46.4276} {\bibfield  {journal}
  {\bibinfo  {journal} {Jap. J. App. Phys.}\ }\textbf {\bibinfo {volume}
  {46}},\ \bibinfo {pages} {4276} (\bibinfo {year} {2007})}\BibitemShut
  {NoStop}%
\bibitem [{\citenamefont {Gueroult}\ \emph {et~al.}(2016)\citenamefont
  {Gueroult}, \citenamefont {Evans}, \citenamefont {Zweben}, \citenamefont
  {Fisch},\ and\ \citenamefont {Levinton}}]{Gueroult2016a}%
  \BibitemOpen
  \bibfield  {author} {\bibinfo {author} {\bibfnamefont {R.}~\bibnamefont
  {Gueroult}}, \bibinfo {author} {\bibfnamefont {E.~S.}\ \bibnamefont {Evans}},
  \bibinfo {author} {\bibfnamefont {S.~J.}\ \bibnamefont {Zweben}}, \bibinfo
  {author} {\bibfnamefont {N.~J.}\ \bibnamefont {Fisch}}, \ and\ \bibinfo
  {author} {\bibfnamefont {F.}~\bibnamefont {Levinton}},\ }\href {\doibase
  10.1088/0963-0252/25/3/035024} {\bibfield  {journal} {\bibinfo  {journal}
  {Plasma Sources Sci. Technol.}\ }\textbf {\bibinfo {volume} {25}},\ \bibinfo
  {pages} {035024} (\bibinfo {year} {2016})}\BibitemShut {NoStop}%
\bibitem [{\citenamefont {Tsushima}\ \emph {et~al.}(1986)\citenamefont
  {Tsushima}, \citenamefont {Mieno}, \citenamefont {Oertl}, \citenamefont
  {Hatakeyama},\ and\ \citenamefont {Sato}}]{Tsushima1986}%
  \BibitemOpen
  \bibfield  {author} {\bibinfo {author} {\bibfnamefont {A.}~\bibnamefont
  {Tsushima}}, \bibinfo {author} {\bibfnamefont {T.}~\bibnamefont {Mieno}},
  \bibinfo {author} {\bibfnamefont {M.}~\bibnamefont {Oertl}}, \bibinfo
  {author} {\bibfnamefont {R.}~\bibnamefont {Hatakeyama}}, \ and\ \bibinfo
  {author} {\bibfnamefont {N.}~\bibnamefont {Sato}},\ }\href {\doibase
  10.1103/PhysRevLett.56.1815} {\bibfield  {journal} {\bibinfo  {journal}
  {Phys. Rev. Lett.}\ }\textbf {\bibinfo {volume} {56}},\ \bibinfo {pages}
  {1815} (\bibinfo {year} {1986})}\BibitemShut {NoStop}%
\bibitem [{\citenamefont {Tsushima}\ and\ \citenamefont
  {Sato}(1991)}]{Tsushima1991}%
  \BibitemOpen
  \bibfield  {author} {\bibinfo {author} {\bibfnamefont {A.}~\bibnamefont
  {Tsushima}}\ and\ \bibinfo {author} {\bibfnamefont {N.}~\bibnamefont
  {Sato}},\ }\bibfield  {booktitle} {\emph {\bibinfo {booktitle} {Journal of
  the Physical Society of Japan}},\ }\href {\doibase 10.1143/JPSJ.60.2665}
  {\bibfield  {journal} {\bibinfo  {journal} {J. Phys. Soc. Jpn.}\ }\textbf
  {\bibinfo {volume} {60}},\ \bibinfo {pages} {2665} (\bibinfo {year}
  {1991})}\BibitemShut {NoStop}%
\bibitem [{\citenamefont {Light}\ \emph {et~al.}(2001)\citenamefont {Light},
  \citenamefont {Chen},\ and\ \citenamefont {Colestock}}]{Light2001}%
  \BibitemOpen
  \bibfield  {author} {\bibinfo {author} {\bibfnamefont {M.}~\bibnamefont
  {Light}}, \bibinfo {author} {\bibfnamefont {F.~F.}\ \bibnamefont {Chen}}, \
  and\ \bibinfo {author} {\bibfnamefont {P.~L.}\ \bibnamefont {Colestock}},\
  }\href {\doibase 10.1063/1.1403415} {\bibfield  {journal} {\bibinfo
  {journal} {Phys. Plasmas}\ }\textbf {\bibinfo {volume} {8}},\ \bibinfo
  {pages} {4675} (\bibinfo {year} {2001})}\BibitemShut {NoStop}%
\bibitem [{\citenamefont {Kent}\ \emph {et~al.}(1969)\citenamefont {Kent},
  \citenamefont {Jen},\ and\ \citenamefont {Chen}}]{Kent1969}%
  \BibitemOpen
  \bibfield  {author} {\bibinfo {author} {\bibfnamefont {G.~I.}\ \bibnamefont
  {Kent}}, \bibinfo {author} {\bibfnamefont {N.~C.}\ \bibnamefont {Jen}}, \
  and\ \bibinfo {author} {\bibfnamefont {F.~F.}\ \bibnamefont {Chen}},\ }\href
  {\doibase 10.1063/1.1692323} {\bibfield  {journal} {\bibinfo  {journal}
  {Phys. Fluids}\ }\textbf {\bibinfo {volume} {12}},\ \bibinfo {pages} {2140}
  (\bibinfo {year} {1969})}\BibitemShut {NoStop}%
\end{thebibliography}
\end{document}